\documentclass[sn-mathphys]{sn-jnl}

\jyear{2021}%

\theoremstyle{thmstyleone}%
\theoremstyle{thmstyletwo}%

\theoremstyle{thmstylethree}%
\usepackage{xcolor}
\newcommand{\sech}{\mbox{sech}}
\newcommand{\Del}{\Delta}
\newcommand{\del}{\delta}

\newcommand{\ba}{\beta}

\newcommand{\ga}{\gamma}

\newcommand{\bea}{\begin{eqnarray}}
	\newcommand{\eea}{\end{eqnarray}}
\newcommand{\bes}{\begin{subequations}}
	\newcommand{\ees}{\end{subequations}}

\begin{document}
	
	\title[Soliton and breather solutions of the generalized LSRI system ]{Bright, dark and breather soliton solutions of the generalized long-wave short-wave resonance interaction system }
	
	
	\author*[1]{\fnm{M.} \sur{Kirane}}\email{mokhtar.kirane@ku.ac.ae; mokhtar.kirane@yahoo.com}
	\author[1]{\fnm{S.} \sur{Stalin}}\email{stalin.seenimuthu@ku.ac.ae}
	
	\author[2]{\fnm{M.} \sur{Lakshmanan}}\email{lakshman@cnld.bdu.ac.in}

	\affil*[1]{\orgdiv{Department of Mathematics, College of Art and Sciences}, \orgname{Khalifa University of Science and Technology}, \orgaddress{\city{Abu Dhabi}, \postcode{127788}, \country{United Arab Emirates}}}
	
	\affil[2]{\orgdiv{Department of Nonlinear Dynamics}, \orgname{Bharathidasan University}, \orgaddress{\street{} \city{Tiruchirappalli}, \postcode{620024}, \state{Tamilnadu}, \country{India}}}
	
	
	
	\abstract{ In this paper,  a generalized long-wave short-wave resonance interaction system, which describes the nonlinear interaction between a short-wave and a long-wave in fluid dynamics, plasma physics and nonlinear optics, is considered. Using the Hirota bilinear method, the general $N$-bright and $N$-dark soliton solutions are deduced and their Gram determinant forms are obtained. A special feature of the fundamental bright soliton solution is that, in general, it behaves like the Korteweg-deVries soliton. However, under a special condition, it also behaves akin to the  nonlinear Schr\"{o}dinger soliton when it loses the amplitude dependent velocity property. The fundamental dark-soliton solution admits anti-dark, grey, and completely black soliton profiles, in the short-wave component, depending on the choice of wave parameters. On the other hand, a bright soliton like profile always occurs in the long-wave component. The asymptotic analysis shows that both the bright and dark solitons undergo an elastic collision with a finite phase shift. In addition to these, by tuning the phase shift regime, we point out the existence of resonance interactions among the bright solitons. Furthermore, under a special velocity resonance condition, we bring out the various types of bright and dark soliton bound states. Also, by fixing the phase factor and the system parameter $\beta$, corresponding to the interaction between long and short wave components, the different types of profiles associated with the obtained breather solution are demonstrated.

	}

	
	
	
	\maketitle
	
	\section{Introduction}\label{sec1}
	Resonance, a nonlinear phenomenon, which often occurs when the wave numbers or frequencies of two or more waves satisfy appropriate resonance condition \cite{ml}. Such a unique physical phenomenon is widely observed in both linear and nonlinear dynamical systems. Among many, a classical example is the long-wave short-wave resonance interaction (LSRI) model which finds applications in fluid dynamics \cite{benny,grimshaw1}, plasma physics \cite{zakharov,oikawa}, nonlinear optics \cite{kivol,lsrinim,ablowitz2}, Bose-Einstein condensation \cite{frantz,frantz1},  and biophysics \cite{boiti}. This LSRI takes place when both high frequency short-wave (SW) and low frequency long-wave (LW) obey the Zakharov-Benney condition: the group velocity of the SW ($v_g=d\omega(k)/ dk$) must exactly or almost matches the phase velocity of the LW ($v_p=\omega/k$).  That is $v_g=v_p$. The LSRI literature originally starts from the theoretical investigation on Langmuir waves in plasma where the generalized Zakharov equations were derived \cite{zakharov}. After this pioneering work by Zakharov, there have been several experimental and theoretical research activities based on the LSRI phenomenon in different contexts ranging from
	lower dimensions \cite{rede,ma,rede1}  to higher dimensions \cite{boyd,funakoshi, ohta1,radha}, with single component \cite{kanna1} to multi-component \cite{kanna2,kanna3,kanna4,chen1,sazonov,jetp2009,myrza86}. These studies also report the existence of several types of nonlinear localized wave structures \cite{kanna1,kanna2, kanna3, chen3,wazwaz,alrazi1,alrazi2,alrazi3}, namely  bright soliton with a single-hump structure \cite{kanna1,kanna2, kanna3, chen3}  and bright soliton with a double-hump structure \cite{stalin-lsri}, dark soliton \cite{chen1,kanna3,kanna4}, breathers \cite{chen5}, and rogue-waves \cite{chow1, crespo1,chow2,crespo2,chen4},
	and their novel properties have also been exhibited there. The main focus of this paper is to present the soliton, both bright and dark, solutions and breather solution for the recently introduced  generalized long-wave short-wave resonance interaction (LSRI) system 
	\begin{eqnarray}
		&&	iS_t+S_{xx}+(i\alpha L_x+\alpha^2L^2-\beta L- 2 \alpha \lvert S\rvert^2)S=0,\nonumber \\
		&&	L_t=2(\lvert S\rvert^2)_x.  \label{1}
	\end{eqnarray}
	The system (\ref{1}) has been introduced in \cite{bib1}, where the authors have established the integrability of the above system by providing its ($3\times 3$) Lax Pair.  In  system (\ref{1}), $S$ $(\equiv S(x,t))$ describes the short-wave and $L$ $(\equiv L(x,t))$ represents the long-wave, and suffices $x$ and $t$ denote the partial
	derivatives with respect to spatial and evolutional coordinates, respectively, and the nonlinearity coefficients $\alpha$ and $\beta$ are real parameters. The nonlinearities arise in Eq. (\ref{1}) because of the self interaction of the short-wave packet, as in the case of NLS equation, and the interaction between LW and SW. The formation of soliton in the SW component is essentially due to the balance of its dispersion by the  nonlinear interactions of LW and SW and the self interaction of the SW. The self interaction of the SW determines the formation and evolution of soliton in the LW component.     
	
	We wish to point out that the generalized LSRI system (\ref{1}) reduces to two well known LSRI models. For example, the system (\ref{1}) becomes the following Yajima-Oikawa (YO for short) system for $\beta=\pm 1$ and $\alpha=0$ \cite{oikawa}, 
	\begin{eqnarray}
		iS_t+S_{xx}\pm LS=0, ~~~
		L_t=2(\lvert S\rvert^2)_x,  \label{2.a}
	\end{eqnarray}
	and it turns into the Newell LSRI system,
	\begin{eqnarray}
		iS_t+S_{xx}+(iL_x+L^2- 2 \lvert S\rvert^2)S=0,~~~
		L_t=2(\lvert S\rvert^2)_x, \label{2.b}
	\end{eqnarray}
	for $\beta=0$ and $\alpha=1$ \cite{newell,ling,bib3}. In Ref. \cite{oikawa}, the formation and the interaction of solitons are studied within the framework of the YO-system (\ref{2.a}) by the inverse scattering technique (IST) while the Langmuir waves coupled with ion-acoustic waves propagating in one-direction. An alternate long-wave short-wave model (\ref{2.b}) has been proposed, and the nature of the solitons is analyzed using IST, by Newell in Ref. \cite{newell} to describe Benney's theory of the nonlinear interaction of long and short waves. The present LSRI system (\ref{1}) proposed in \cite{bib1} can be treated as the general one to explain the interaction of long and  short-waves.
	
	From the literature, we find that the nature of the solitons,  their underlying analytical forms and their interaction properties have not been unravelled so far for Eq. (\ref{1}). This is what it is intended to be reported in this paper. By applying the Hirota bilinear method, multi-bright and multi-dark-soliton solutions of the system (1) are constructed along with the breather solution. An important fact is that these multi-soliton solutions are written in a compact way using the Gram determinants. By doing so, we find that the fundamental bright soliton of the present LSRI system behaves like the KdV soliton since it possesses the amplitude dependent velocity property. While imposing a special condition on the system parameter $\beta$ and the velocity of soliton, it also acts like the NLS soliton. The existence of these properties simultaneously in the present generalized LSRI system (\ref{1}) is not possible in the other single and multi-component YO LSRI systems \cite{oikawa,bib3,kanna1} and in the derivative YO or Newell LSRI system too \cite{newell,bib3}. Further, very interestingly, the bright solitons undergo V and Y-type resonance interactions by tuning the phase shift regime. Such a possibility is not observed earlier in the YO-system (\ref{2.a}). In addition to these, an interesting fact which we observe in the present LSRI system is the appearance of a standing breather in the breather patterns. We get the soliton in a periodic wave pattern by tuning the background wave field.   Apart from these, by fixing the velocity resonance condition appropriately, various types of bright and dark bound states are also brought out. 
	
	In general, to solve any integrable nonlinear partial differential equations (PDEs), the following analytical methods have been widely used in the soliton literature \cite{ml,istbook,dtbook}. For instance, (i). Inverse scattering transform, (ii). Darboux transformation method, (iii). B\"{a}cklund transformation method, (iv). Hirota bilinear method and (v). Lie-symmetry analysis. The first four methods have been used to derive more general soliton solutions, whereas using the last Lie symmetry analysis a limited class of solitary wave solutions/similarity solutions can be derived by reducing the given nonlinear PDE into an ordinary differential equation.  Each of the methods have their own advantages and demerits. One can derive all possible soliton solutions, including breathers, rogue waves, bright, and dark soliton solutions, using the above first four methods. However, it is not possible to derive such solutions using the Lie symmetry analysis, which only provides the information about the solitary wave solutions not the general soliton solutions. We also briefly point out these various aspects in Table I.
	
	\begin{table}[h]
			\begin{minipage}{174pt}
				\caption{Advantages and disadvantages of the various analytical methods}\label{tab1}%
				\begin{tabular}{@{}llll@{}}
					\toprule
					Method  & Advantage & Disadvantage\\
					\midrule
					 Inverse Scattering Transform   & \multirow{3}{9em}{Multi-soliton solutions can be obtained and the Cauchy initial value problem can be solved completely } & Too technical \vspace{2.0cm} \\
					Darboux transformation   & \multirow{3}{9em}{Multi-soliton solutions can be obtained }& \multirow{3}{9em}{Cauchy initial value problem cannot be solved fully}  \vspace{0.8cm}\\
					 B\"{a}cklund transformation   & \multirow{3}{9em}{Multi-soliton solutions can be obtained }  & \multirow{3}{9em}{Cauchy initial value problem cannot be solved fully}  \vspace{0.6cm}\\
					\\
					 Hirota bilinear method   &\multirow{3}{9em}{Multi-soliton solutions can be obtained }& \multirow{3}{9em}{Cauchy initial value problem cannot be solved fully}  \vspace{0.6cm}  \\
					\\
					Lie-symmetry analysis   & \multirow{3}{9em}{Solitary wave solutions/similarity solutions can be obtained}  & \multirow{3}{9em}{Only particular solutions can be obtained}  \vspace{1.0cm}\\
					\botrule
				\end{tabular}\end{minipage}
		\end{table}
	
	The rest of the paper is organized as follows: In Sect. 2, the fundamental as well as the higher-order bright soliton solutions are derived, and the various interaction  dynamics associated with the bright solitons are explained in Sect. 3 with appropriate asymptotic analysis. The one-and two-dark-soliton solutions are given in Sect. 4 and the various possible collision dynamics of two-dark solitons are explained in Sect. 5. In Sect. 6, we demonstrate the breather solution of the system (\ref{1}) and its characteristics with suitable graphical illustration. In Sect. 7, the obtained results are summarized. For completeness, the $N$-bright and $N$-dark soliton solutions in Appendices A and $B$, respectively, are presented.     
	\section{Bright soliton solutions}
	To derive the soliton and breather solutions of the system (\ref{1}), the Hirota bilinear method, in which  one has to introduce an appropriate bilinearizing transformation in order to obtain the bilinear forms of a given nonlinear partial differential equation, is adopted. Following  Hirota \cite{hirota}, to get the bilinear forms of Eq. (\ref{1}), we introduce the bilinearizing  transformations
	\begin{eqnarray}
		S(x,t)=\frac{g}{f},~ ~L(x,t)=i\frac{\partial}{\partial x}\log\frac{f^*}{f}, ~~g\equiv g(x,t),~~f\equiv f(x,t),\label{2}
	\end{eqnarray}
	in it. In the above, both the unknown functions $g$ and $f$ are complex functions. While doing the bilinearization of Eq. (\ref{1}), we choose  $\alpha=1$, without loss of generality. Substitution of (\ref{2}) in Eq. (\ref{1}) yields its corresponding bilinear forms as given below: 
	\begin{eqnarray}
		(iD_t+D_x^2)g\cdot f=0, ~~ i(D_t+\beta D_x)f\cdot f^*=D_x^2f\cdot f^*,~~iD_t f\cdot f^*=-2gg^*, \label{3}
	\end{eqnarray}
	where the Hirota's bilinear operators $D_x$ and $D_t$ are defined in \cite{hirota}. 
	Substituting the standard expansions for the unknown functions $g$ and $f$,
	\begin{eqnarray}
		g=\epsilon g_1+\epsilon^3 g_3+...,~~~~
		f=1+\epsilon^2 f_2+\epsilon^4 f_4+...,
		\label{4}
	\end{eqnarray}
	in Eqs. (\ref{3}), one gets a system of linear PDEs. The set of linear PDEs arises after collecting the coefficients of same powers of  $\epsilon$, which is a formal series expansion parameter, and equating the terms corresponding to each power of $\epsilon$ individually to zero. By solving these linear PDEs recursively (at an appropriate order of $\epsilon$), we obtain the explicit forms of $g$ and $f$. Such explicit forms constitute the bright soliton solutions to the underlying generalized LSRI system (\ref{1}).
	\begin{figure}[]
		\centering
		\includegraphics[width=0.85\linewidth]{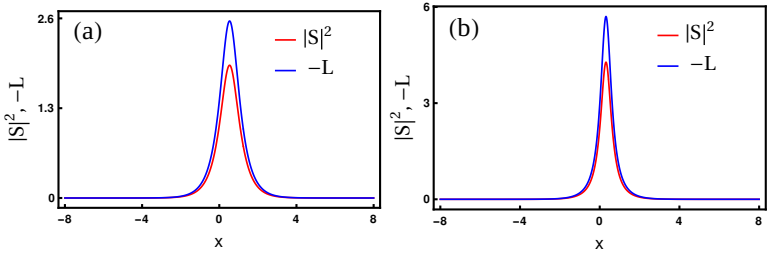}
		\caption{Fundamental bright soliton of the generalized LSRI system (\ref{f1}) is illustrated in Fig. (a) for $k_1=1+0.75i$, $\ga_1=1$, and $\beta=1$. The corresponding soliton compression graph is depicted in Fig. (b) for the system parameter $\beta=-1$. }
		\label{f1}
	\end{figure}
	
	\subsection{One-soliton solution}
	The fundamental bright soliton solution of the system (\ref{1}) can be obtained by solving the following set of equations
	\bes
	\bea
	&&D_1g_1\cdot 1=0, ~~~D_2(1\cdot f_2^*+f_2\cdot 1)=D_x^2(1\cdot f_2^*+f_2\cdot 1)\nonumber\\
	&&iD_t(1\cdot f_2^*+f_2\cdot 1)=-2g_1g_1^*,\nonumber
	\eea
	\ees
	along with the initial seed solution, $g_1=\gamma_1e^{\eta_1}$, $\eta_1=k_1x+ik_1^2t$. 
	Here, $D_1$ and $D_2$ are defined as $D_1\equiv iD_t+D_x^2$, $D_2\equiv i(\beta D_x+D_t)$, respectively. The explicit forms of $g_1$ and $f_2$ give rise to  the fundamental bright soliton solution of the system (\ref{1}). It reads as 
	\bes\begin{eqnarray}
		&&S(x,t)=\frac{\epsilon g_1 }{1+\epsilon^2 f_2}=\frac{\ga_1e^{\eta_1}}{1+e^{\eta_1+\eta_1^*+\delta}}, \label{5a}\\
		&&L(x,t)=i\frac{\partial }{\partial x}\log\frac{1+\epsilon^2 f_2^*}{1+\epsilon^2 f_2}=i\frac{\partial }{\partial x} \log\frac{1+e^{\eta_1+\eta_1^*+\delta^*}}{1+e^{\eta_1+\eta_1^*+\delta}},\label{5b}
	\end{eqnarray}\ees
	where  
	$\displaystyle{e^{\del}=  \frac{\lvert\ga_1\rvert^2(i\ba+2k_1^*)}{(k_1+k_1^*)^2(k_1-k_1^*)}}$. The small parameter $\epsilon$ does not contribute anything to the structure of soliton and so one can choose it as $1$, without loss of generality (or subsume as an additional constant in the wave variable $\eta_1$).  The profile structures of the SW and LW are described by the two complex constants $k_1$ and $\ga_1$ and the system parameter $\ba$. We wish to note that the bright soliton solution (\ref{5a})-(\ref{5b}) exactly coincides with the already reported fundamental bright soliton solution of the derivative LSRI system \cite{bib3} when $\ba=0$. Therefore, the fundamental bright soliton solution derived by us for model (\ref{1}) can be considered as more general. To understand the properties of the obtained soliton solution (\ref{5a})-(\ref{5b}) further, we rewrite it in hyperbolic form. It turns out to be    
	\bes\begin{eqnarray}
		&&S(x,t)=A_Se^{i\eta_{1I}}\sech(\eta_{1R}+\frac{\delta}{2}),~~~A_S=k_{1R}\bigg(\frac{2\ga_1k_{1I}}{\ga_1^*(\beta-2ik_1^*)}\bigg)^{\frac{1}{2}},\label{6a}\\ &&L(x,t)=\frac{A_L}{\frac{(\beta-2k_{1I})}{\lvert 2k_1-i\beta\rvert}+\cosh(2\eta_{1R}+\frac{\delta+\delta^*}{2})}, ~~~A_L=-\frac{4k_{1R}^2}{\lvert 2k_1-i\beta\rvert},\label{6b}
	\end{eqnarray}\ees
	where $\eta_{1R}=k_{1R}(x-2k_{1I}t)$, $\eta_{1I}=k_{1I}x+(k_{1R}^2-k_{1I}^2)t$. Here, $\eta_{1R}$, $k_{1R}$, and $\eta_{1I}$, $k_{1I}$ are the real and imaginary parts of $\eta_1$, and $k_1$, respectively. In the above, $A_S$ and $A_L$ represent the respective amplitudes of the soliton in the SW and LW components and they propagate from $-x$ to $+x$ direction with the velocity $v=2k_{1I}$. Note that $\frac{\delta}{2}=\frac{1}{2}\log \frac{\lvert\ga_1\rvert^2(i\ba+2k_1^*)}{(k_1+k_1^*)^2(k_1-k_1^*)}$ is complex. The central positions of the SW and LW are obtained as  $\frac{\delta+\delta^*}{4k_{1R}}=-\frac{1}{4k_{1R}}\log\frac{\lvert\ga_1\rvert^4(i\ba+2k_1^*)(i\ba-2k_1)}{(k_1+k_1^*)^4(k_1-k_1^*)^2}$. A typical profile of the fundamental bright soliton solution of the system (\ref{1}) is displayed in Fig. \ref{f1}(a). Then, we plot the solution (\ref{6a})-(\ref{6b}) in Fig. \ref{f1} (b) with $\beta<0$. The graph clearly demonstrates that the soliton profiles, in both the SW and LW components, are compressed significantly. This kind of simultaneous amplification and  compression of optical pulses is indeed observed in an experiment \cite{agrawal} and it is useful in nonlinear optics applications to generate picosecond or femtosecond pulses.
	\begin{figure}[]
		\centering
		\includegraphics[width=0.4\linewidth]{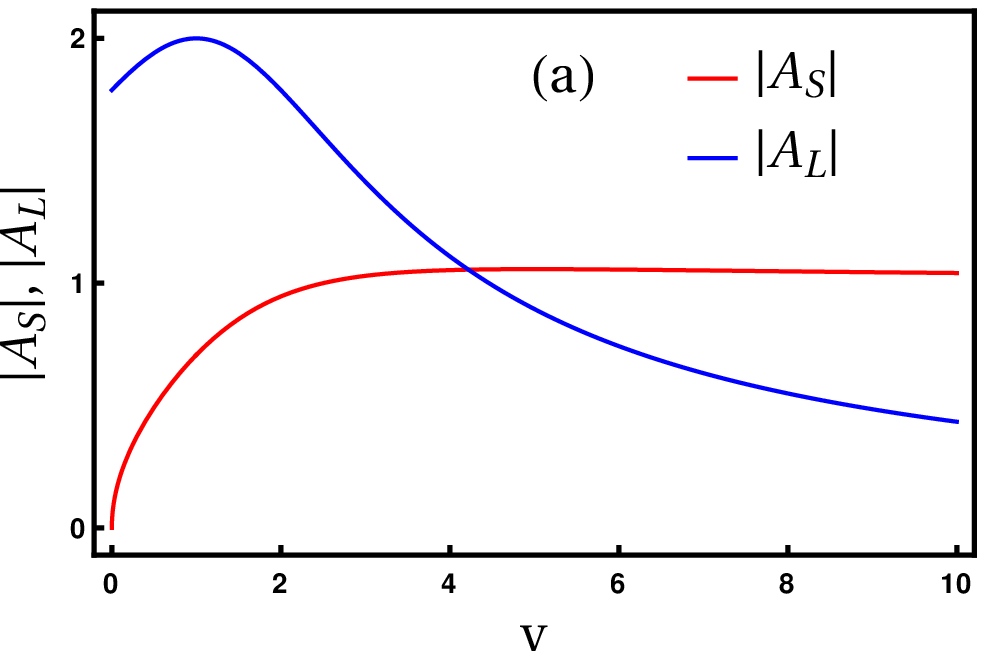}~	\includegraphics[width=0.4\linewidth]{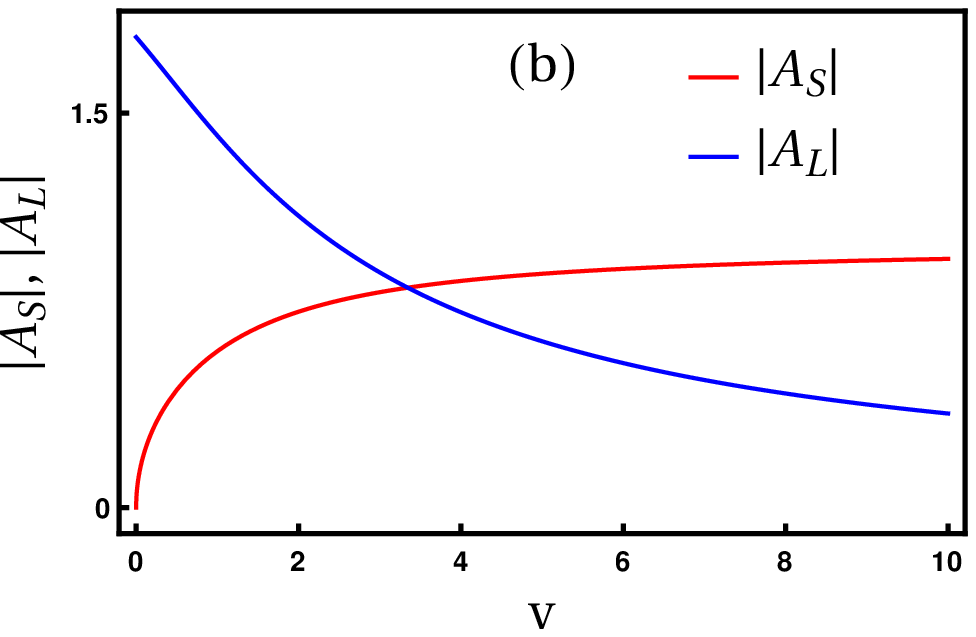}\\
		\includegraphics[width=0.4\linewidth]{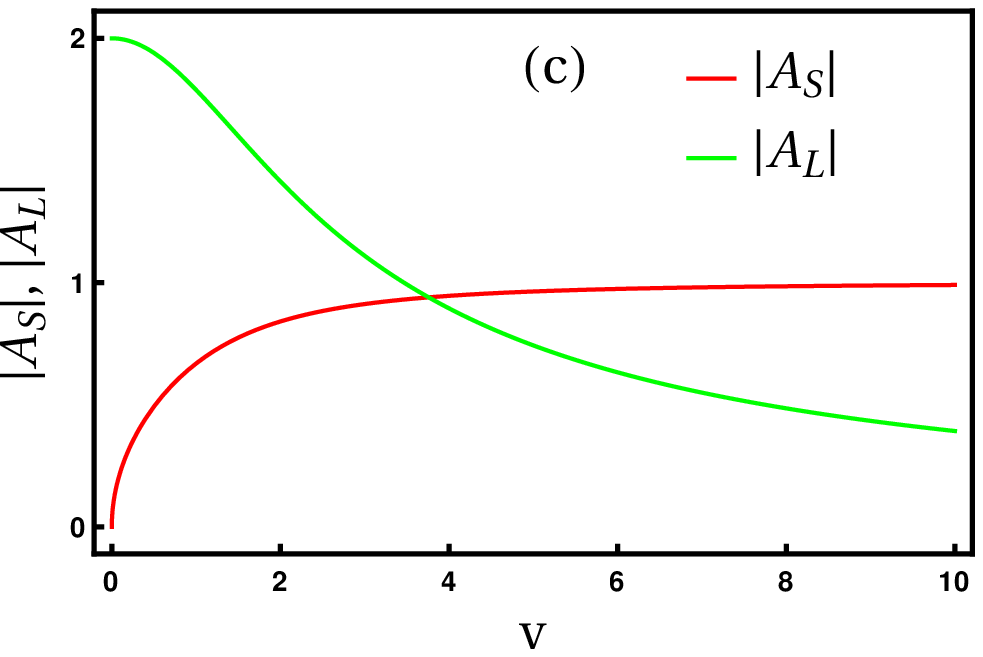}~	\includegraphics[width=0.4\linewidth]{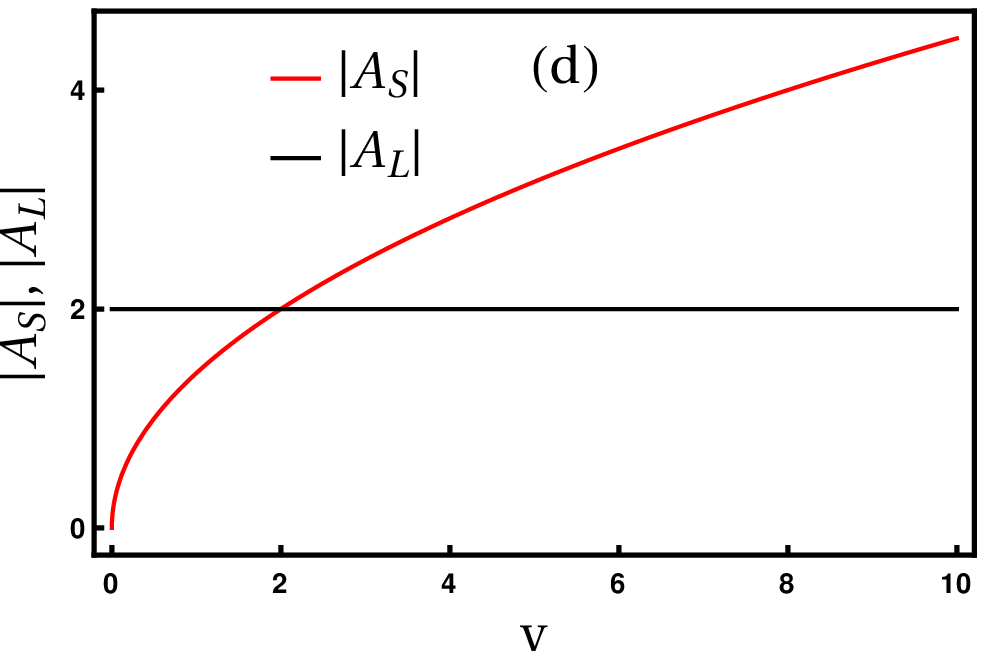}
		\caption{Amplitude-velocity relation of the fundamental bright soliton in the present LSRI system (\ref{1}) and in the other LSRI models (\ref{2.a}) and (\ref{2.b}). To draw Fig. \ref{f2}(a) we fix the parameter values as $k_{1R}=0.5$, $\beta=1$,  and $\ga_{1}=1$. For Figs. \ref{f2}(b), and \ref{f2}(c) we consider $\beta$ value as $-1$, $\beta=0$, respectively and the other values remain the same as in the previous case. For   Fig. \ref{f2} (d) we illustrate the amplitude-velocity relation graph for the  YO system with $k_{1R}=0.5$,  and $\ga_{1}=1$.   }
		\label{f2}
	\end{figure}
	
	Another interesting property associated with the fundamental bright soliton solution (\ref{6a})-(\ref{6b}) of the system (\ref{1}) is the explicit appearance of soliton velocity in the amplitude parts of both the SW and LW components. As a result, the taller soliton will travel faster not only in the SW component, but also in the LW component. This special property is akin to KdV solitons \cite{ml}. We remark that this interesting property is distinct from the property of fundamental bright soliton of the YO system \cite{oikawa}, where the velocity appears only in the SW component. Such amplitude dependent velocity property of the bright soliton is illustrated in Fig. \ref{f2} for the present generalized LSRI system (\ref{1}) and the other single component LSRI systems (\ref{2.a}) and (\ref{2.b}) \cite{bib3,oikawa}. For instance, in the present LSRI system (\ref{1}) with $\beta>0$, we find that the amplitude $A_L$ is decreasing with respect to $v$ whereas the amplitude $A_S$ is increasing as illustrated in Fig. \ref{f2}(a). We also observe a similar scenario for $\beta<0$, which is illustrated in Fig. \ref{f2}(b). Further, for completeness, we draw the amplitude-velocity relation graphs for the other cases, the derivative YO system ($\beta=0$) \cite{bib3}, and the YO system $\alpha=0$, $\beta=-1$ \cite{oikawa,ma}, in Figs. \ref{f2}(c) and (d), respectively. 
	
	It is very important to point out that the bright soliton in the  generalized system also acts like the NLS bright soliton for the choice $\beta=2k_{1I}$. That is the soliton in the underlying system (\ref{1}) no longer possesses the  amplitude dependent velocity  property. In this situation, the solution (\ref{6a})-(\ref{6b}) gets reduced as 
	\bes\bea
	&& S(x,t)=(\frac{i\ga_1}{\ga_1^*})^{1/2}k_{1R}e^{i\eta_{1I}}\sech(\eta_{1R}+\frac{\delta}{2}),~~~\frac{\delta}{2}=\frac{1}{2}\log\frac{\lvert\ga_1\rvert^2}{\sqrt{4ik_{1R}^2k_{1I}}},~\label{nls-sol1}\\
	&&L(x,t)=-2k_{1R}\sech(2\eta_{1R}+\frac{\delta+\delta^*}{2}),~~~\frac{\delta+\delta^*}{2}=\frac{1}{2}\log\frac{\lvert\ga_1\rvert^4}{16k_{1R}^4k_{1I}^2}.\label{nls-sol2}
	\eea\ees
	The latter expressions clearly indicate that the amplitude of the soliton does not depend on its velocity and the bright soliton of the form (\ref{nls-sol1})-(\ref{nls-sol2}) propagates like the NLS bright soliton  with the velocity $2k_{1I}$. This interesting property is not possible in the other single and multi-component LSRI systems \cite{oikawa,bib3,kanna1}.       
	\subsection{Two-soliton solution}
	Next, we find that the two series forms in Eq. (\ref{4}) get terminated for the two bright soliton solution of the system (\ref{1}) as  $g=\epsilon g_1+\epsilon^3 g_3$ and $f=1+\epsilon^2 f_2+\epsilon^4 f_4$. The resultant forms constitute the two-soliton solution and it turns out to be 
	\bes
	\begin{eqnarray}
		&&\hspace{-0.5cm}S=\frac{1}{f}\bigg(\ga_1e^{\eta_1}+\ga_2e^{\eta_2}+\Delta_{121^*}e^{\eta_1+\eta_2+\eta_1^*}+\Delta_{122^*}e^{\eta_1+\eta_2+\eta_2^*}\bigg),\label{7a}\\
		&&\hspace{-0.5cm}L=i\frac{\partial}{\partial x}\log\frac{f^*}{f},\\
		&&\hspace{-0.5cm}f=1+\delta_{11^*}e^{\eta_1+\eta_1^*}+\delta_{12^*}e^{\eta_1+\eta_2^*}+\delta_{21^*}e^{\eta_2+\eta_1^*}+\delta_{22^*}e^{\eta_2+\eta_2^*}\nonumber \\&&\hspace{0.2cm} +\delta_{121^*2^*}e^{\eta_1+\eta_1^*+\eta_2+\eta_2^*},\label{7c}\\
		&&\hspace{-0.5cm}\delta_{ij^*}=\frac{\ga_{i}\ga_{j}^*(i\beta+2k_j^*)}{(k_i+k_j^*)^2(k_i-k_j^*)},~\Delta_{12i^*}=(k_2-k_1)\big(\frac{\ga_2\delta_{1i^*}}{k_2+k_i^*}-\frac{\ga_1\delta_{2i^*}}{k_1+k_i^*}\big),\nonumber\\
		&&\hspace{-0.5cm}\delta_{121^*2^*}=\vert k_1-k_2\rvert^2\bigg[\frac{\delta_{11^*}\delta_{22^*}}{(k_1+k_2^*)(k_2+k_1^*)}-\frac{\delta_{12^*}\delta_{21^*}}{(k_1+k_1^*)(k_2+k_2^*)}\bigg]~i,j=1,2,\nonumber\label{7d} 
	\end{eqnarray}\ees
	where $\eta_j=k_jx+ik_j^2t$, $j=1,2$. The above two-soliton solution is characterized by four-arbitrary complex parameters, $k_j$ and $\ga_j$, $j=1,2$ and one system parameter $\beta$. These parameters non-trivially contribute to the collision properties of the two bright solitons as we explain below. We also get the explicit forms of $N$-bright soliton solution of the generalized LSRI system, which is given in Appendix \ref{secA1}.    
	\section{Collision dynamics of bright solitons}
	The interesting aspect of the generalized LSRI system (\ref{f1}) is that the bright solitons associated with it undergo different types of interactions apart from the standard elastic collision. For example, they exhibit (i) resonance interactions and (ii) soliton bound state or soliton molecule for the appropriate choices of wave parameters. First, we perform the asymptotic analysis in order to confirm the elastic nature of collision among the two bright solitons, then we will analyse the resonance interactions and soliton bound states in detail.     
	\subsection{Elastic collision: Asymptotic analysis }
	To study the interaction dynamics of the solitons completely, we perform a detailed asymptotic analysis of the two-soliton solution (\ref{7a})-(\ref{7c}) and deduce the explicit forms of the individual solitons at the limits $t\rightarrow \pm\infty$. To investigate this, we consider
	$k_{jR}>0$, $j=1,2$, $k_{1I}>k_{2I}$, which corresponds to either the case of a head-on collision or the case of an overtaking collision between the two solitons (depending on the signs of $k_{jI}$'s). However, here, we have considered the head-on collision among the two bright solitons. In this situation the two fundamental solitons are well separated and subsequently the asymptotic forms of the individual solitons can be deduced from the solution (\ref{7a})-(\ref{7c}) by incorporating the asymptotic nature of the wave variables $\eta_{jR}=k_{jR}(x-2k_{jI}t)$, $j=1,2$, in it. The wave variables $\eta_{jR}$'s behave asymptotically as (i) Soliton 1: $\eta_{1R}\simeq 0$, $\eta_{2R}\rightarrow\pm \infty$ as $t\rightarrow \pm\infty$ and (ii) Soliton 2: $\eta_{2R}\simeq 0$, $\eta_{1R}\rightarrow\mp \infty$ as $t\rightarrow\pm\infty$. Correspondingly, these results lead to the following asymptotic forms of individual bright solitons.\\
	(a) Before collision: $t\rightarrow -\infty$\\
	Soliton 1: In this limit, the asymptotic forms of both the SW and LW are deduced from the two-soliton solution (\ref{7a})-(\ref{7c}) for soliton 1 as given below: 
	\bes\bea
	&&\hspace{-0.5cm}S(x,t)\simeq A_S^{1-}e^{i\eta_{1I}}\sech(\eta_{1R}+\phi_S^{1-}),~~~A_S^{1-}=k_{1R}\bigg(\frac{2\ga_1k_{1I}}{\ga_1^*(\beta-2ik_1^*)}\bigg)^\frac{1}{2},\label{8a}\\ 
	&&\hspace{-0.5cm}L(x,t)\simeq \frac{A_L^{1-}}{\frac{(\beta-2k_{1I})}{\lvert2k_1-i\beta\rvert}+\cosh(2\eta_{1R}+\phi_L^{1-})},~~~ A_L^{1-}=-\frac{4k_{1R}^2}{\lvert2k_1-i\beta\rvert},\label{8b}
	\eea\ees
	where the phase terms are given by
	\bea
	\phi_S^{-1}=\frac{1}{2}\log\frac{\lvert\ga_1\rvert^2(i\beta+2k_1^*)}{(k_1+k_1^*)^2(k_1-k_1^*)},~~~\phi_L^{-1}=\frac{1}{2}\log\frac{-\lvert\ga_1\rvert^4\lvert 2k_1-i\beta\rvert^2}{(k_1+k_1^*)^4(k_1-k_1^*)^2}.\nonumber
	\eea
	In the latter, superscript ($1-$) represents the soliton $1$ before collision and the suffices $S$ and $L$ denote the SW and LW components, respectively.\\
	Soliton 2: The following asymptotic forms of the soliton 2 are deduced from the solution (\ref{7a})-(\ref{7c}). They read as
	\bes\bea
	&&\hspace{-0.8cm}S(x,t)\simeq A_S^{2-}e^{i(\eta_{2I}+\theta_2)}\sech(\eta_{2R}+\phi_S^{2-}),~A_S^{2-}=k_{2R}\bigg(\frac{2\ga_2k_{2I}}{\ga_2^*(\beta-2ik_2^*)}\bigg)^\frac{1}{2},\label{9a}\\ &&\hspace{-0.8cm}L(x,t)\simeq\frac{A_L^{2-}}{\frac{(2k_{2I}-\beta)}{\lvert2k_2-i\beta\rvert}+\cosh(2\eta_{2R}+\phi_L^{2-})},~ A_L^{2-}=\frac{4k_{2R}^2}{\lvert 2k_2-i\beta\rvert},\label{9b}\\
	&&\hspace{-0.8cm}e^{i\theta_2}=\frac{(k_1-k_2)(k_1+k_2^*)(k_1+k_2)^{\frac{1}{2}}(k_2^*-k_1)^{\frac{1}{2}}}{(k_1^*-k_2^*)(k_1^*+k_2)(k_1^*+k_2^*)^{\frac{1}{2}}(k_2-k_1^*)^{\frac{1}{2}}}.\nonumber
	\eea\ees  
	Here, the phase terms are defined as
	\bea &&\phi_S^{2-}=\frac{1}{2}\log\frac{\lvert\ga_2\rvert^2(i\beta+2k_2^*)\lvert k_1-k_2\rvert^4\lvert k_1+k_2\rvert^2}{\lvert k_1-k_2^*\rvert ^2\lvert k_1+k_2^*\rvert^4(k_2-k_2^*)(k_2+k_2^*)^2},\nonumber \\\text{and}&&
	\phi_L^{2-}=\frac{1}{2}\log\frac{\lvert \ga_2\rvert^4(i\beta+2k_2^*)(i\beta-2k_2)\lvert k_1-k_2\rvert^8\lvert k_1+k_2\rvert^4}{\lvert k_1-k_2^*\rvert^4\lvert k_1+k_2^*\rvert^8(k_2-k_2^*)^2(k_2+k_2^*)^4}.\nonumber\eea
	In the latter, superscript ($2-$) represents the soliton $2$ before collision.  \\
	(b) After collision: $t\rightarrow +\infty$\\
	Soliton 1: Similarly, in this long time limit, the asymptotic forms of both the SW and LW are obtained as
	\bes\bea
	&&\hspace{-0.8cm}S(x,t)\simeq A_S^{1+}e^{i(\eta_{1I}+\theta_1)}\sech(\eta_{1R}+\phi_S^{1+}),~A_S^{1+}=k_{1R}\bigg(\frac{2\ga_1k_{1I}}{\ga_1^*(\beta+2ik_1^*)}\bigg)^\frac{1}{2},\label{10a}\\ &&\hspace{-0.8cm}L(x,t)\simeq\frac{A_L^{1+}}{\frac{(\beta-2k_{1I})}{\lvert 2k_1-i\beta\rvert}+\cosh(2\eta_{1R}+\phi_L^{1+})},~~ A_L^{1+}=-\frac{4k_{1R}^2}{\lvert2k_1-i\beta\rvert},\label{10b}\\
	&&\hspace{-0.8cm}e^{i\theta_1}=\frac{(k_1-k_2)(k_1+k_2)^{\frac{1}{2}}(k_1^*-k_2)^{\frac{1}{2}}}{(k_1^*-k_2^*)(k_1^*+k_2^*)^{\frac{1}{2}}(k_1-k_2^*)^{\frac{1}{2}}}. \nonumber
	\eea\ees
	The corresponding phase terms are calculated as
	\bea
	&&\phi_S^{1+}=\frac{1}{2}\log\frac{\lvert\ga_1\rvert^2\lvert k_1-k_2\rvert^4\lvert k_1+k_2\rvert^2(2k_1^*+i\beta)}{(k_1-k_1^*)(k_1+k_1^*)^2\lvert k_1-k_2^*\rvert^2\lvert k_1+k_2^*\rvert^4},\nonumber \\
	\text{and}&& \phi_L^{1+}=\frac{1}{2}\log\frac{\lvert\ga_1\rvert^4\lvert k_1-k_2\rvert^8\lvert k_1+k_2\rvert^4(2k_1^*+i\beta)(-2k_1+i\beta)}{(k_1-k_1^*)^2(k_1+k_1^*)^4\lvert k_1-k_2^*\rvert^4\lvert k_1+k_2^*\rvert^8}.\nonumber \eea
	In the the latter, superscript ($1+$) represents the soliton $1$ after collision.\\
	Soliton 2: For the soliton 2, the asymptotic expressions turn out to be
	\bes\bea
	&&S(x,t)\simeq A_S^{2+}e^{i\eta_{2I}}\sech(\eta_{2R}+\phi_S^{2+}),~A_S^{2+}=k_{2R}(\frac{2\alpha_2k_{2I}}{\alpha_2^*(\beta-2ik_2^*)})^\frac{1}{2},\label{}\\ &&L(x,t)\simeq\frac{A_L^{2+}}{\frac{(\beta-2k_{2I})}{\lvert2k_2-i\beta\rvert }+\cosh(2\eta_{2R}+\phi_L^{2+})},~ A_L^{2+}=-\frac{4k_{2R}^2}{\lvert2k_2-i\beta\rvert},\label{}
	\eea\ees
	where
	\bea &&\phi_S^{2+}=\frac{1}{2}\log\frac{\lvert\ga_2\rvert^2(i\beta+2k_2^*)}{(k_2+k_2^*)^2(k_2-k_2^*)}, \nonumber\\
	\text{and} && \phi_L^{2+}=\frac{1}{2}\log\frac{\lvert\ga_2\rvert^4(i\beta+2k_2^*)(i\beta-2k_2)}{(k_2+k_2^*)^4(k_2-k_2^*)^2}.\nonumber \eea
	\begin{figure*}[]
		\centering
		\includegraphics[width=0.5\linewidth]{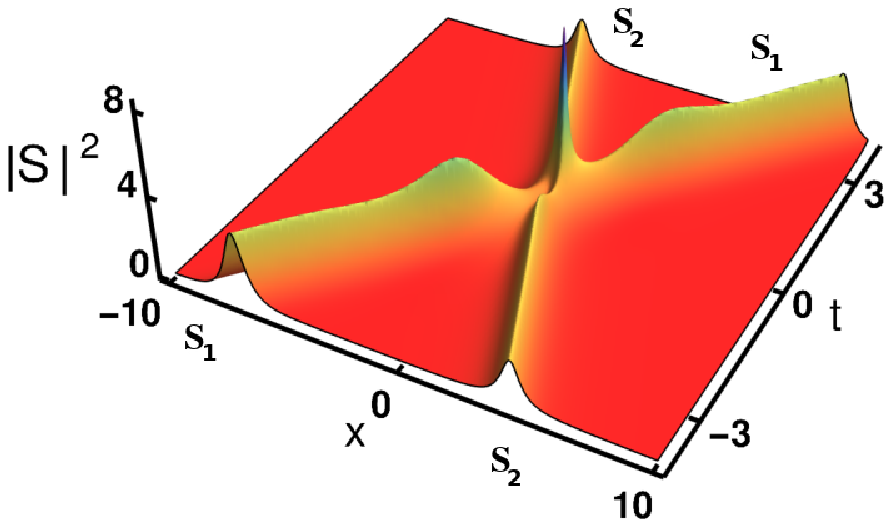}~	\includegraphics[width=0.5\linewidth]{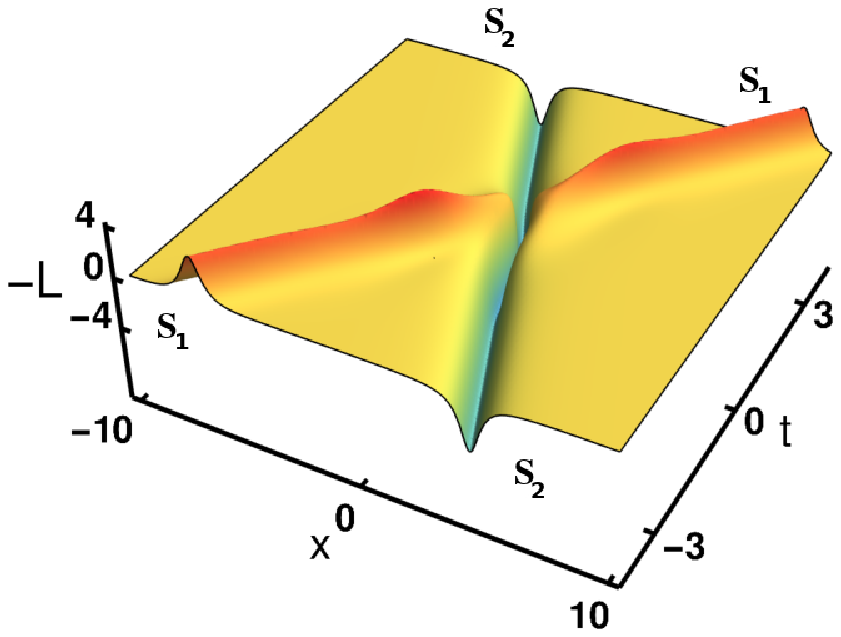}
		\caption{Elastic collision among the two bright solitons of the system (\ref{1}). The parameter values are $k_1=1+i$, $k_2=0.5-0.5i$, $\ga_1=0.8$, $\ga_2=0.45$ and $\beta=1$. }
		\label{f3}
	\end{figure*}
	\begin{figure*}[]
		\centering
		\includegraphics[width=0.8\linewidth]{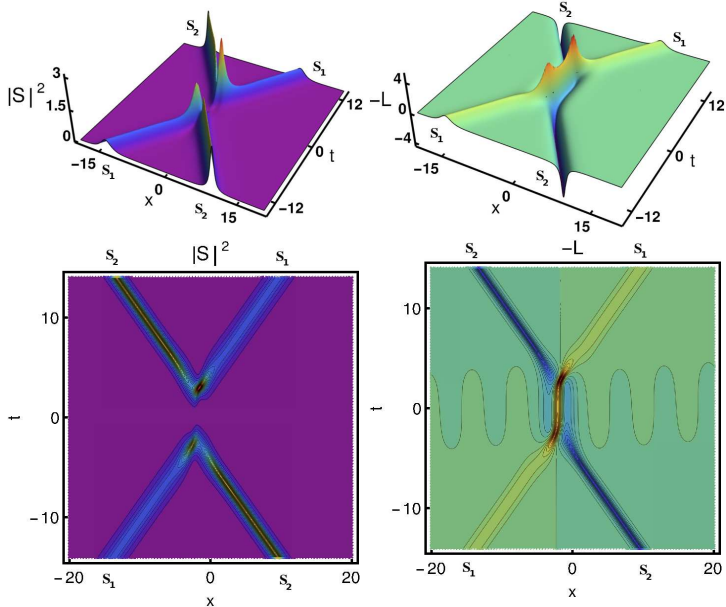}
		\caption{In the top panel, the resonance interaction among the two bright solitons are demonstrated and the corresponding space-time plots are given in the bottom panel. They shows that the two bright solitons take a finite time to interact in both the SW and LW components.   }
		\label{f4}
	\end{figure*}
	\begin{figure*}[]
		\centering
		\includegraphics[width=0.8\linewidth]{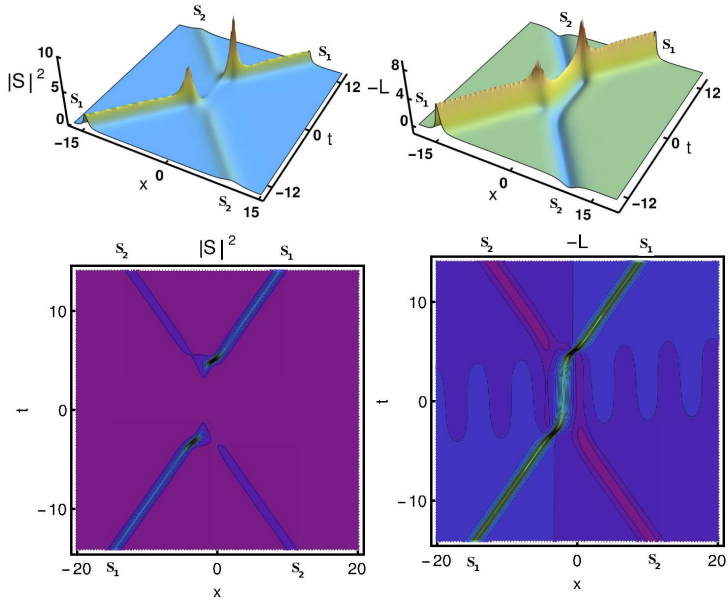}
		\caption{In the top panel, the resonance interaction among the two bright solitons is illustrated and the corresponding space-time plot is demonstrated in the bottom panel. Here, the resonance interaction happens, among the two bright solitons, for a longer time period than one in Fig. \ref{f4}. This is achieved by tuning the phase shift regime further.  }
		\label{f5}
	\end{figure*}
	\begin{figure*}[]
		\centering
		\includegraphics[width=0.4\linewidth]{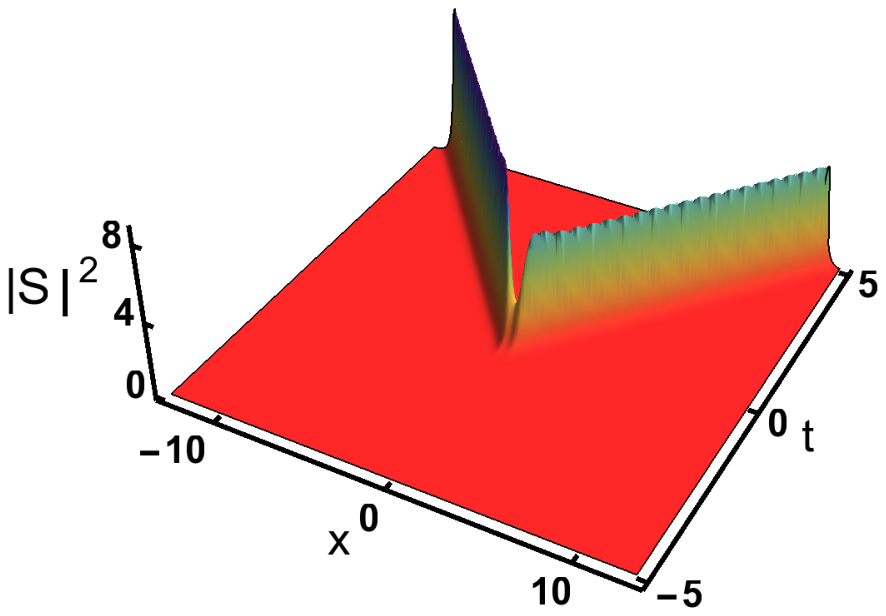}~	\includegraphics[width=0.4\linewidth]{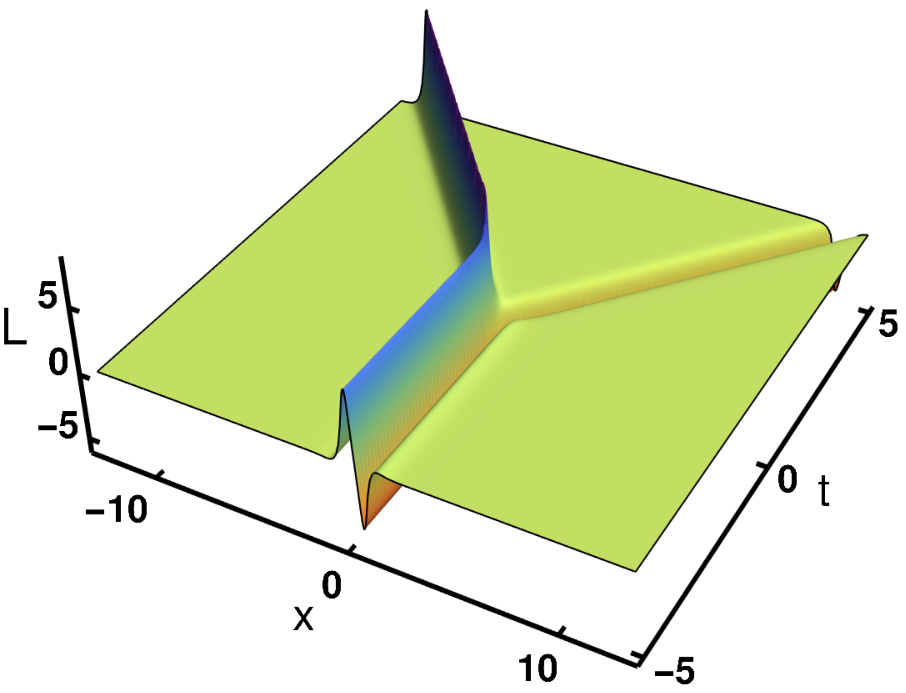}
		\caption{V and Y type resonance interactions among the two bright solitons. They arise by setting the phase shift $\Delta \Phi_S=\Delta \Phi_L\rightarrow \infty$. It can be fixed by setting the condition $k_{2R}=-k_{1R}$ and $k_{2I}=-k_{1I}$.}
		\label{f6}
	\end{figure*}
	\begin{figure*}[]
		\centering
		\includegraphics[width=0.4\linewidth]{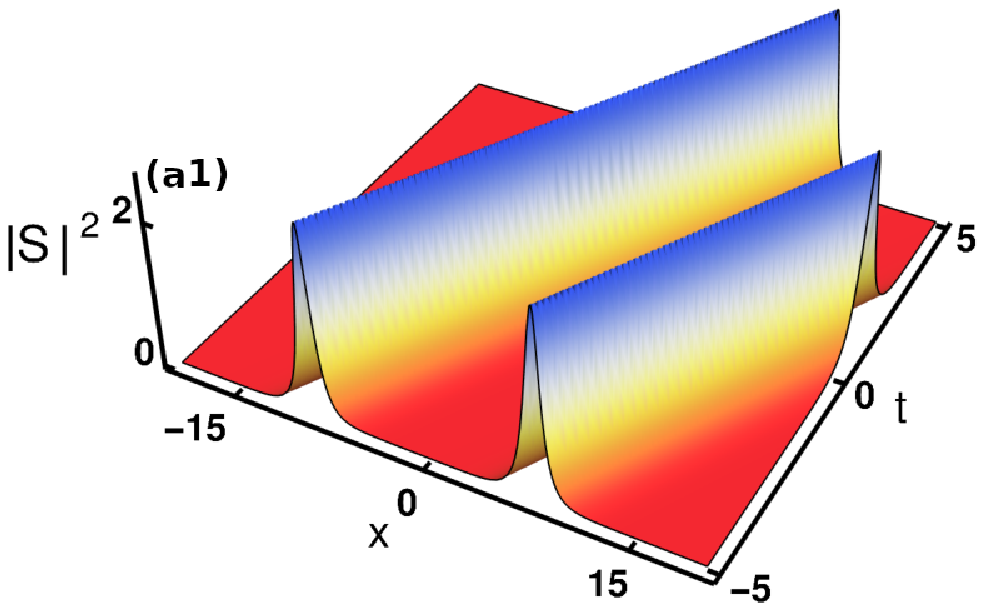}~	\includegraphics[width=0.4\linewidth]{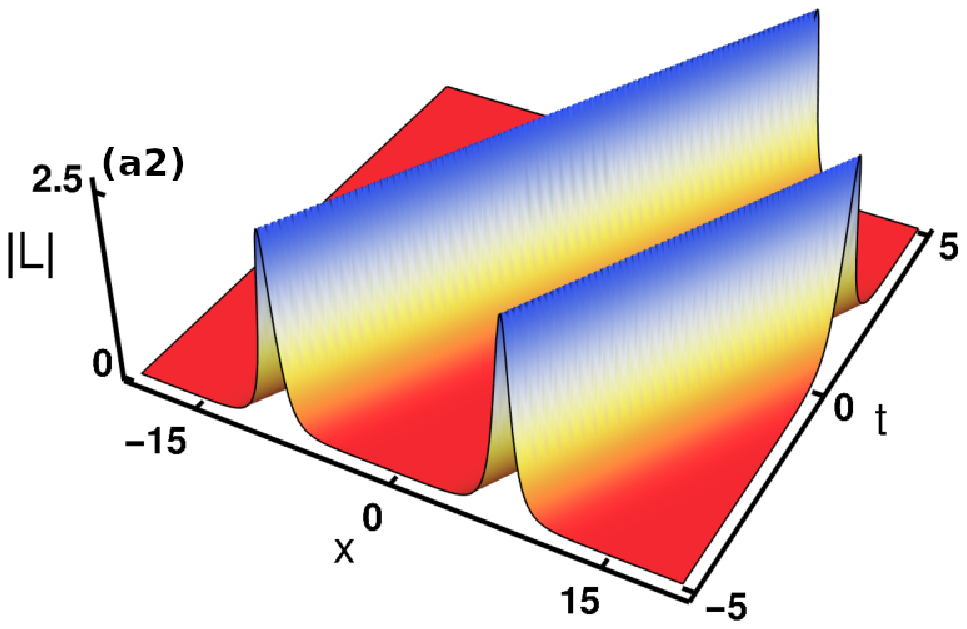}\\
		\includegraphics[width=0.4\linewidth]{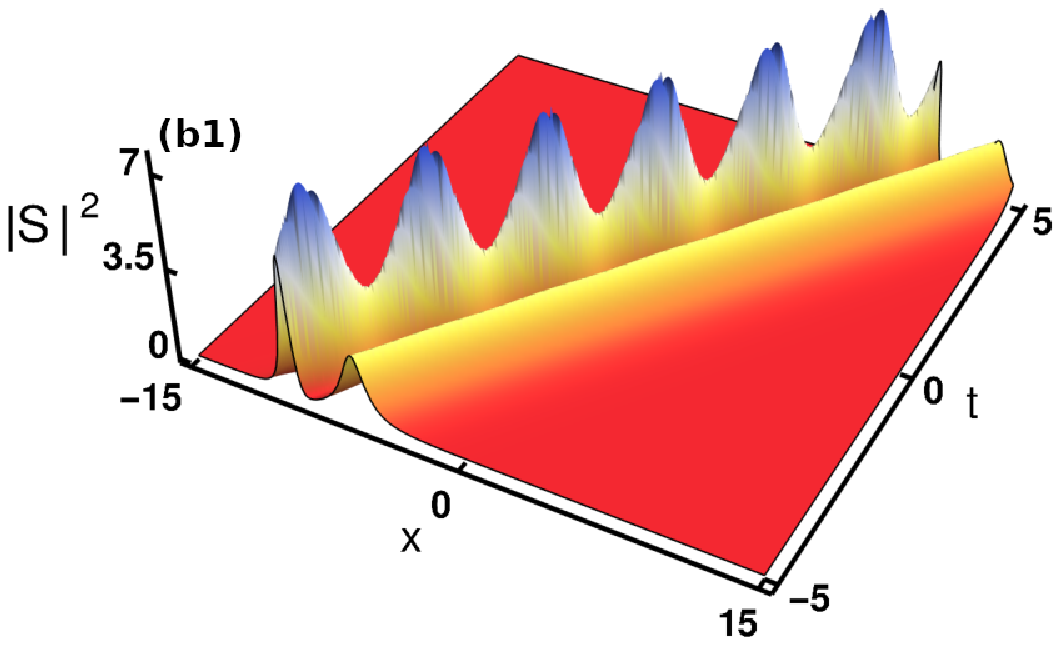}~	\includegraphics[width=0.4\linewidth]{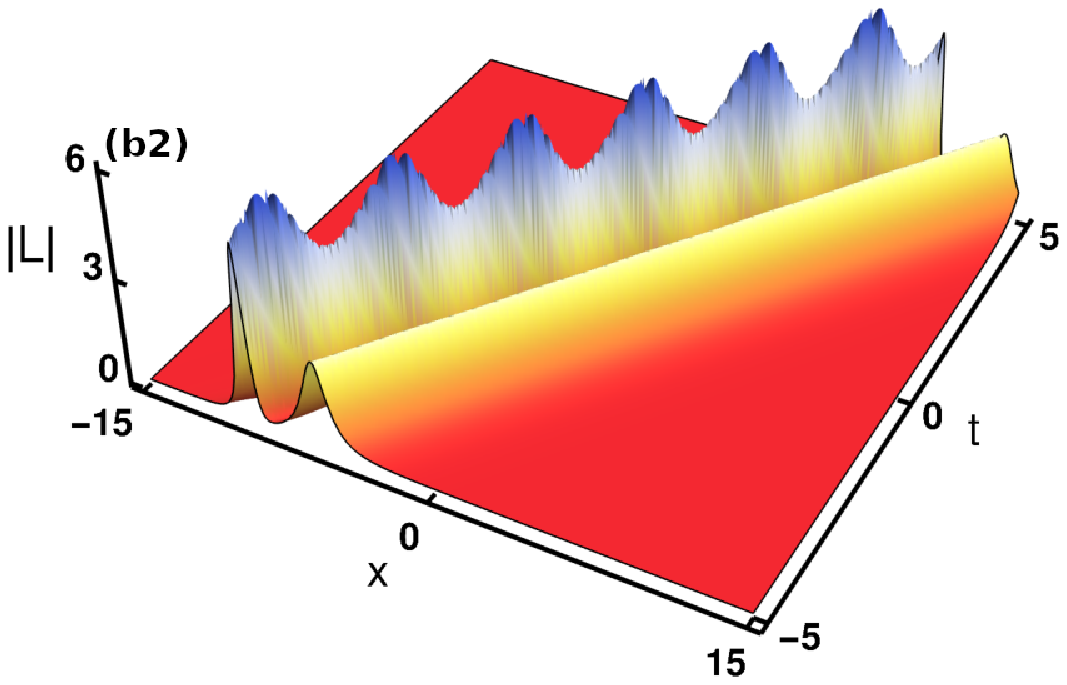}\\
		\includegraphics[width=0.4\linewidth]{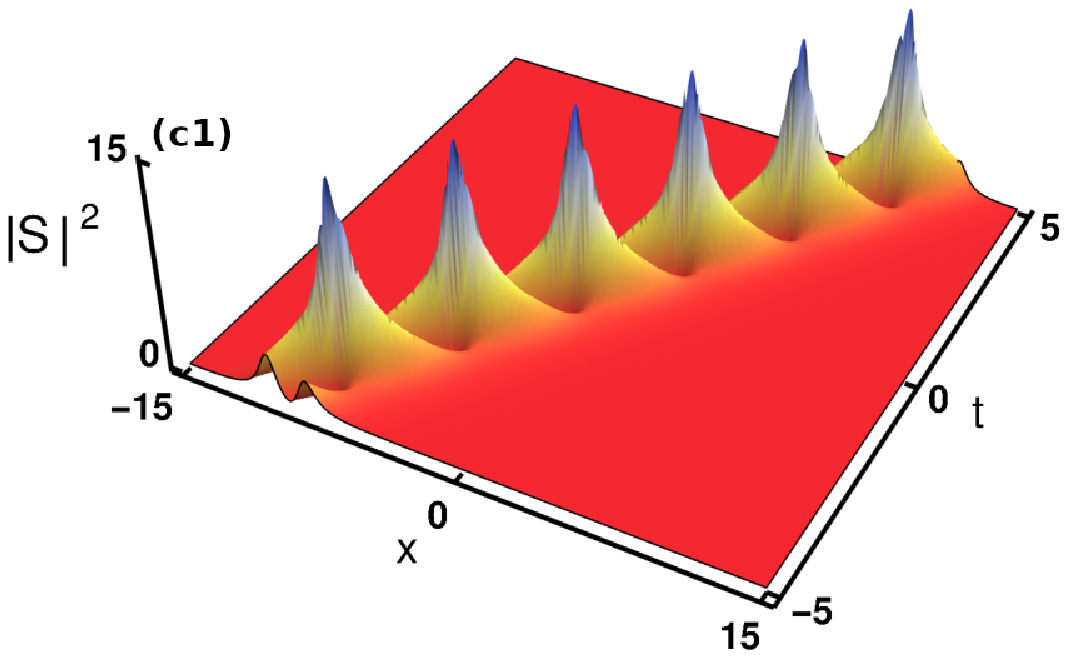}~	\includegraphics[width=0.4\linewidth]{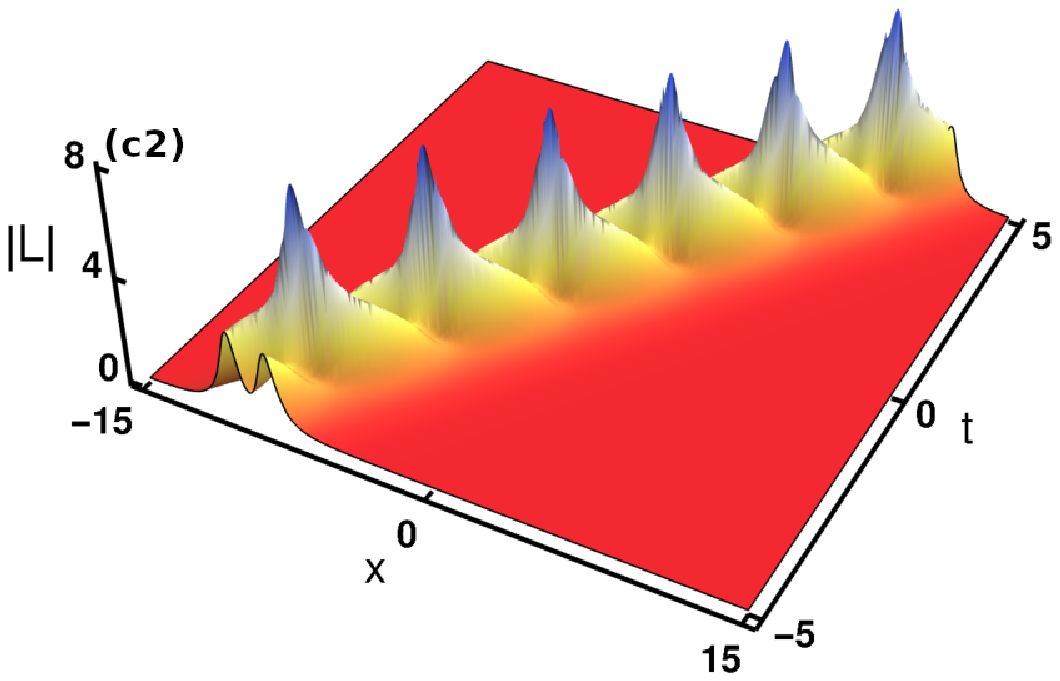}\\
		\caption{Top panel denotes the parallel propagation bound soliton state. Middle panel represents breathing-type bound state where oscillation occurs in one of the solitons and bottom panel illustrates breathing type-soliton state where oscillations occur in both the solitons. }
		\label{f7}
	\end{figure*}
	The above asymptotic analysis shows that the amplitudes of the solitons remain the same before and after collisions. Consequently,  the transition intensities are always unimodular. That is, \bea \lvert T_S^{j}\rvert^2=\frac{\lvert A_S^{j+}\rvert^2}{\lvert A_S^{j-}\rvert^2}=1, ~\text{and}~
	\lvert T_L^{j}\rvert^2=\frac{\lvert A_L^{j+}\rvert^2}{\lvert A_L^{j-}\rvert^2}=1, ~j=1,2.\eea
	It implies that the bright solitons of the generalized LSRI system always undergo a shape preserving collision, with a finite phase shift, thereby confirming the elastic nature of the collision. Correspondingly, the energy of each of the solitons is conserved.  Such an elastic collision is displayed in Fig. \ref{f3}, where the dark-like profile appears in the LW component essentially because of the negative sign that arises in the amplitude part. The phase shifts suffered by the solitons in both the SW and LW components are obtained as
	\begin{equation}
		\Del\Phi_S^1=\frac{1}{2}\log\frac{\lvert k_1-k_2\rvert ^4\lvert k_1+k_2\rvert^2}{\lvert k_1+k_2^*\rvert ^4\lvert k_1-k_2^*\rvert^2}=-\Del\Phi_S^2,~\Del\Phi_L^1=-\Del\Phi_L^2=2\Phi_S^1.\label{13}
	\end{equation} 
	The above implies that the two bright solitons are located exactly opposite with each other after the collision process and their positions are mainly influenced by the wave numbers, $k_j$, $j=1,2$.   
	\subsection{Resonance interactions}
	The bright solitons of the system (\ref{1}) exhibit interesting resonance interaction patterns for appropriately chosen wave parameters. These patterns will appear in the interaction regime during the soliton collision and they can be viewed as an intermediate state. Such a state  essentially arises when the phase shifts due to collision become infinity or larger value. A typical example of resonance interaction pattern is depicted in Fig. \ref{f4} for the parameter values $k_1=0.5+0.5i$, $k_2=0.45-0.5i$, $\ga_1=0.9$, $\ga_2=0.45$. The figure shows that the interaction regime 
	gets extended and the two solitons take larger time to interact. This is clearly distinct from the standard collision, which is demonstrated in Fig. \ref{f3}, where the interaction happens without much delay. However, in Fig. \ref{f4} the interaction period is finite and after that the two bright solitons split and travel with their own velocities. One can realize that the intermediate state in the SW component as zero amplitude soliton as discussed in the case of the higher dimensional LSRI system \cite{kanna3}. In contrast to this, we observe a standing breather, like pattern that appears in the LW component. Such pattern exists only for a shorter duration and it is clearly different from the one that has been widely discussed in the rogue wave theory. We wish to note that one can also tune the interaction regime further by setting a condition  $k_{2R}\approx k_{1R}$ along with the choice $\beta=-1$. This is illustrated in Fig. \ref{f5}. 
	
	Apart from the above pattern, we  also observed another interesting interaction pattern when we fix the condition $k_{2R}=-k_{1R}$ and $k_{2I}=-k_{1I}$. We call such a pattern as a V-Y type resonance interaction pattern which is displayed in Fig. \ref{f6} with the parameter values $k_1=2+1.05i$, $k_2=-2-1.05i$, $\ga_1=0.25$, $\ga_2=0.5$, and $\beta=1$. From this figure, we observe that the interaction regime becomes infinity. This is because of the fact that the phase shifts $\Del \Phi_S$ and $\Del \Phi_L$ (Eq. (\ref{13})) tend to infinity for $k_2=-k_1$.  Consequently, in the SW component, the two bright solitons approach each other only asymptotically and form a zero amplitude resonant soliton, whereas in the LW component they form a standing breather pattern which is extended up to an infinite interaction regime. 
	
	Next, we show that the existence of different types of bound soliton state or soliton molecule, which is recently a hot topic in soliton theory and has potential applications in optical telecommunications. This novel structure essentially arises when the two solitons propagate with either equal or nearly equal velocity and it can be considered as a special case of the standard two soliton solution (interacting soliton state). Depending on the choice of the central position, there exist two types of such soliton state: (i) Parallel propagation, and (ii) Breather. We find that these bound soliton structures also exist  in the generalized LSRI system (\ref{1}). To explore the bound soliton state in Eq. (\ref{f1}), we fix the velocity resonance condition as $v_1(=2k_{1I})\approx v_2(=2k_{2I})$ and also $k_{1R}=k_{2R}$ so that the two bright solitons can propagate with almost the same velocity and they form a soliton molecule structure.  A typical  parallel propagating bound soliton state is displayed in Fig. \ref{f7}(a1)-(a2). To obtain this soliton state we fix the parameter values as $\beta=1$, $k_1=0.65+i$, $k_2=0.65+0.99i$, $\ga_1=0.5$ and $\ga_2=0.35$. Then to get the breathing soliton molecule, we consider the same velocity resonance condition but with $k_{1R}\neq k_{2R}$. The outcome is depicted in Fig. \ref{f7}(c1)-(c2), where the two bright solitons exhibit oscillatory behaviors. By fixing the parameter values as $\beta=1$, $k_1=2+i$, $k_2=0.5+0.995i$, $\ga_1=1$ and $\ga_2=1.35$, we bring out this soliton molecule structure. This breathing soliton molecular structure can be easily identified by rewriting the two-soliton solution (\ref{7a})-(\ref{7c}) in hyperbolic forms. The resultant forms will contain trigonometric functions $\cos(\eta_{1I}-\eta_{2I})$ and $\sin(\eta_{1I}-\eta_{2I}$) in the denominator of both the expressions for $S(x,t)$ and $L(x,t)$. Due to this fact, breathing behaviour emerges in the bright-soliton bound states. Note that one can tune the oscillatory behavior in any one of the solitons by tuning the values of $\ga_j$'s. For example, we control the oscillation that occurs in the second soliton by fixing $\ga_2=0.35$ and keeping all the other parameters the same as the one used in Fig. \ref{f7}(c1)-(c2). A typical graph of such bound soliton state is illustrated in Fig. \ref{f7}(b1)-(b2). It clearly indicates that the oscillation completely suppressed in the second soliton while it still persists in the other soliton structure.        
	
	\section{Dark soliton solutions}
	Next, to derive the dark-soliton solution, now we consider the following transformations \cite{bib3}
	\begin{eqnarray}
		S(x,t)=\tau e^{i\theta}\frac{g(x,t)}{f(x,t)},~ L(x,t)=i\frac{\partial}{\partial x}\log\frac{f^*}{f},~ \theta=lx-(l^2+2\lvert\tau\rvert^2)t. \label{14}
	\end{eqnarray}
	While deriving the dark-soliton solutions one has to consider the non-vanishing boundary condition $S\rightarrow \tau e^{i\theta}$ and $L\rightarrow 0$ when $\lvert x\rvert \rightarrow \infty$, which are included in the above transformations. 
	Here $\tau$ is a complex constant and $l$ is a real constant. Substituting Eq. (\ref{14}) in Eq. (\ref{1}), we arrive at the bilinear forms of Eq. (\ref{1}). They read as\bes
	\begin{eqnarray}
		&&(iD_t+2ilD_x+D_x^2)g\cdot f=0, ~~ i(D_t+\beta D_x)f\cdot f^*=D_x^2f\cdot f^*,\label{15a}\\
		&&iD_t f\cdot f^*=2\lvert\tau\rvert^2(\lvert f\rvert^2-\lvert g\rvert^2). \label{15b}
	\end{eqnarray}\ees
	By solving these bilinear equations along with series expansions, 
	\begin{equation}
		g(x,t)=1+\epsilon g_1+\epsilon^2 g_2+\epsilon^3 g_3+...,~~ f(x,t)=1+\epsilon f_1+\epsilon^2 f_2+\epsilon^3 f_3+...,
	\end{equation}
	we obtain the fundamental as well as multi-dark soliton solutions as given below. 
	\subsection{One-dark soliton solution}
\begin{figure*}[]
		\centering
		\includegraphics[width=0.85\linewidth]{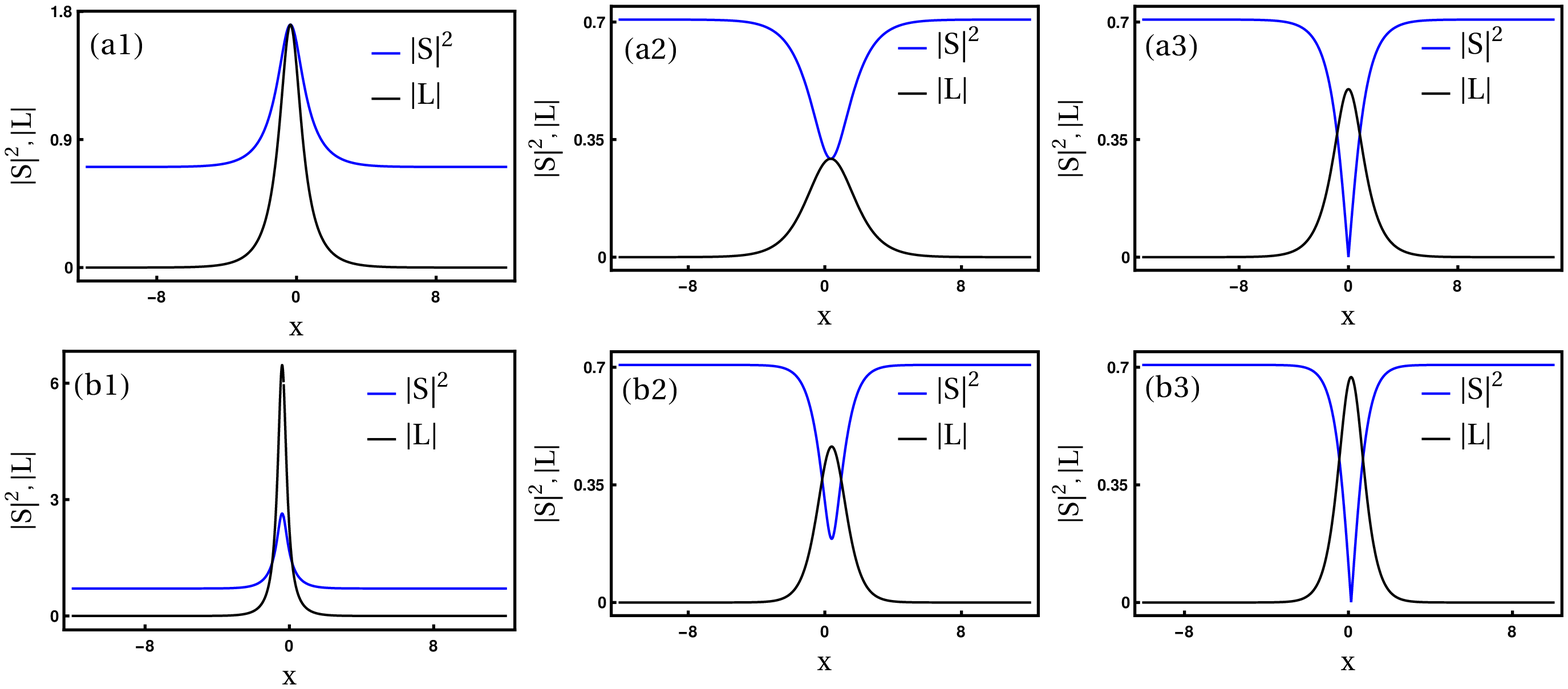}	
		\caption{Various fundamental dark-soliton profiles of the system (\ref{1}) are shown. In Fig. (a1) we depict an anti-dark soliton profile whereas a grey soliton profile is displayed in Fig. (a2). A complete black or dark soliton profile is illustrated in Fig. (a3). In all these figures the corresponding bright soliton profile is drawn in the LW component. The bottom panel (b1)-(b3) displays their corresponding shape compression plots for $\beta=-1$.  }
		\label{f8}
	\end{figure*}

	The fundamental dark soliton solution of the system (\ref{1}) is obtained as 
	\bes\bea
	&&S(x,t)=\tau e^{i\theta}\frac{1+\epsilon g_1}{1+\epsilon f_1}=\tau e^{i\theta}\frac{1+z_1e^{\eta_1+\eta_1^*}}{1+y_1e^{\eta_1+\eta_1^*}},~z_1=-\frac{p_1-il}{p_1^*+il}y_1,\\
	&&L(x,t)=i\frac{\partial}{\partial x}\log\frac{1+y_1^*e^{\eta_1+\eta_1^*}}{1+y_1e^{\eta_1+\eta_1^*}},~y_1=-i\frac{i\beta+2p_1^*}{p_1+p_1^*},
	\eea\ees
	along with a constraint condition \begin{equation}
		p_{1R}=\pm \bigg[\frac{\lvert\tau\rvert^2(2l-\beta)}{2p_{1I}}-(p_{1I}-l)^2\bigg]^{\frac{1}{2}}.\label{18}
	\end{equation}
	Here, $\eta_1=p_1x+ip_1^2t+\eta_{1}^{(0)}$, where $p_1$ and $\eta_1^{(0)}$ are complex constants. 
	The above fundamental dark soliton solution can be rewritten as 
	\bes
	\bea
	&&S(x,t)=\frac{\tau}{2}e^{i\theta}\bigg[(1+\kappa)-(1-\kappa)\tanh(\eta_{1R}+\frac{\delta}{2})\bigg],\label{19a}\\
	&&L(x,t)=- \frac{4p_{1R}^2}{(\beta-2p_{1I})+\lvert 2p_{1R}+i(\beta-2p_{1I})\rvert \cosh(2\eta_{1R}+\frac{\delta+\delta^*}{2})},\label{19b}~
	\eea
	\ees
	where $\kappa=-\frac{p_1-il}{p_1^*+il}$, $e^{\delta}=-i\frac{i\beta+2p_1^*}{p_1+p_1^*}$, and  $\eta_{1R}=p_{1R}(x-2p_{1I}t+\frac{\eta_{1R}^{(0)}}{p_{1R}})$. The dark-soliton solution (\ref{19a})-(\ref{19b}) is described by three complex constants, $\tau$, $p_1$ and $\eta_1^{(0)}$ and two real constants, $l$ and $\beta$. The dark-soliton propagates in both the SW and LW components with the velocity $v=2p_{1I}$. The solution (\ref{19a}) admits an anti-dark soliton on a constant background $\lvert \tau\rvert^2$ in the SW component when $p_{1R}>0$, otherwise it admits dark (or grey) soliton  for $p_{1R}<0$. However, the solution (\ref{19b}) always exhibits bright soliton nature in the LW component. These possibilities are demonstrated in Fig. \ref{f8}. For example, in Fig. \ref{f8}(a1), we display an anti-dark (SW) and bright soliton (LW) profiles for $p_1=1+0.5i$, $\tau=0.5+0.5i$, $l=1$ and $\beta=1$. From this figure, one can observe that an anti-dark soliton is definitely distinct from the usual bright soliton because it appears on a non-vanishing background field.  Then, we illustrate a grey soliton profile in Fig. \ref{f8}(a2), where the intensity of the soliton is lower than the constant background and it does not reach zero intensity anywhere along the $x$-axis. We bring out such a grey soliton profile by fixing the value of $p_{1R}$ as  $-0.5$ and the other parameter values are taken as the same as the one fixed in Fig. \ref{f8} (a1). We also display a dark or black-soliton profile with minimum intensity (intensity reached to zero) in Fig. \ref{f8}(a3) for $p_{1I}=0.5$. In Figs. \ref{f8}(b1)-(b3), we depict their corresponding shape compression plots for $\beta=-1$. We wish to remark that the dark soliton of the generalized LSRI system (\ref{1}) also possesses the amplitude dependent velocity property as the dark soliton has been clearly explained in the case of the derivative YO system \cite{bib3}. 
	
\begin{figure*}[]
		\centering
		\includegraphics[width=0.85\linewidth]{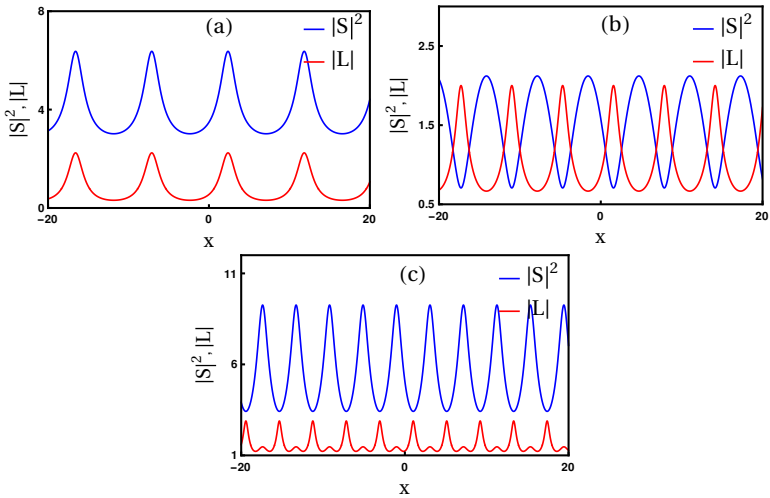}	
		\caption{Periodic solution of the generalized LSRI system (\ref{1}). In Fig. (a), we display in-phase periodic waves whereas anti-phase periodic waves are demonstrated in Fig. (b). A doubly periodic wave is brought out in Fig. (c).   }
		\label{f9}
	\end{figure*}

	Further, interestingly, we also observe that the dark-soliton solution (\ref{19a})-(\ref{19b}) turns into a periodic solution for a lower values of $l$. In this situation, the wave number $p_1$ turns out to be pure imaginary so that hyperbolic form of the dark soliton solution becomes a periodic function. Such a possibility is illustrated in Fig. \ref{f9} with different $l$ values and $\beta>0$. For $l=0.6$, $p_1=1.33i$, $\tau=0.5+0.5i$, and $\beta=1$, we find that in-phase periodic waves appear in both the SW and LW components whereas  anti-phase periodic waves occur for $l=0.5$, $p_1=1.5i$ (the other parameter values are same as the one mentioned above). These examples are displayed in Figs. \ref{f9}(a) and (b), respectively. A doubly-periodic wave arises in the LW component for the choice $l=0.3$ and $p_1=1.76i$. An interesting fact that can be observed from Figs. \ref{f9}(b) and (c) is that in the LW component the intensities of the periodic waves are higher than the background field. This feature is striking contrast with the soliton profiles that are drawn in Fig. \ref{f8}, where all the soliton profiles in the LW component appear only in the zero background. However, we also observe the zero background periodic wave in the LW component. This is demonstrated in Fig. \ref{f9}(a). Note that one can also observe a similar kind of periodic waves in the case of $\beta<0$.    
	\subsection{Two-dark soliton solution}
	The two-dark soliton solution of the generalized LSRI system (\ref{f1}) is derived and it reads as
	\bes\bea
	&&S(x,t)=\tau e^{i\theta}\frac{1+\epsilon g_1+\epsilon^2 g_2}{1+\epsilon f_1+\epsilon^2 f_2}\nonumber\\
	&&\hspace{1.0cm}=\tau e^{i\theta}\frac{1+z_1e^{\eta_1+\eta_1^*}+z_2e^{\eta_2+\eta_2^*}+z_{12}e^{\eta_1+\eta_1^*+\eta_2+\eta_2^*}}{1+y_1e^{\eta_1+\eta_1^*}+y_2e^{\eta_2+\eta_2^*}+y_{12}e^{\eta_1+\eta_1^*+\eta_2+\eta_2^*}},\label{20a}\\
	&&L(x,t)=i\frac{\partial}{\partial x}\log\frac{1+\epsilon f_1^*+\epsilon^2 f_2^*}{1+\epsilon f_1+\epsilon^2 f_2}\nonumber\\
	&&\hspace{1.0cm}=i\frac{\partial}{\partial x}\log\frac{1+y_1^*e^{\eta_1+\eta_1^*}+y_2^*e^{\eta_2+\eta_2^*}+y_{12}^*e^{\eta_1+\eta_1^*+\eta_2+\eta_2^*}}{1+y_1e^{\eta_1+\eta_1^*}+y_2e^{\eta_2+\eta_2^*}+y_{12}e^{\eta_1+\eta_1^*+\eta_2+\eta_2^*}},\label{20b}
	\eea\ees
	where  $\eta_j=p_jx+ip_j^2t+\eta_{j}^{(0)}$, $z_j=-\frac{(p_j-il)}{(p_j^*+il)}y_j$,~ $y_j=-i\frac{(i\beta+2p_j^*)}{(p_j+p_j^*)}$,~$p_{jR}=\pm \bigg[\frac{\lvert\tau\rvert^2(2l-\beta)}{2p_{jI}}-(p_{jI}-l)^2\bigg]^{\frac{1}{2}}$, ~$j=1,2$,~ $z_{12}=z_1z_2\Omega_{12}$, $y_{12}=y_1y_2\Omega_{12}$, $\Omega_{12}=\frac{\lvert p_1-p_2\rvert^2}{\lvert p_1+p_2\rvert^2}$. The two-dark soliton solution (\ref{20a})-(\ref{20b}) is characterized by five complex constants $p_j$, $\eta_{j}^{(0)}$, $j=1,2$, $\tau$ and two real constants $l$ and $\beta$. These parameters control the dynamics as well as the structures of two dark solitons and they also provide the possibility of obtaining three permissible collision scenarios, namely (i) anti-dark - anti-dark solitons collision, (ii) anti-dark - dark solitons collision, and (iii) dark-dark solitons collision. These collision scenarios are analyzed in the subsequent section. We have also obtained $N$-dark soliton solution of the system (\ref{1}), which is given in Appendix B.  
	\section{Collision dynamics of dark solitons: Asymptotic analysis}
	As we have mentioned above, we came across three types of collision scenarios between the dark-solitons. To characterize each of them we have performed the appropriate asymptotic analysis, from which we deduce the explicit forms of the individual dark solitons at the asymptotic time limit $t\rightarrow \pm \infty$. However, here we present the asymptotic analysis corresponding to head-on collision among the two anti-dark solitons only. To perform it we consider the parametric choice, $p_{1R}<p_{2R}$, $p_{1I}>p_{2I}$. By following the procedure described in the case of collision among the bright solitons, we also deduce the following asymptotic forms for anti-dark solitons.\\
	(a) Before collision: $t\rightarrow -\infty$\\
	Soliton 1: $\eta_{1R}\simeq 0$, $\eta_{2R}\rightarrow -\infty$ 
	\bes
	\bea
	&&S(x,t)=\frac{\tau}{2}e^{i\theta}\bigg[(1+\kappa_1)-(1-\kappa_1)\tanh(\eta_{1R}+\phi_{S}^{1-})\bigg],\label{24a}\\
	&&L(x,t)=- \frac{4p_{1R}^2}{(\beta-2p_{1I})+\lvert 2p_{1R}+i(\beta-2p_{1I})\rvert \cosh(2\eta_{1R}+\phi_{L}^{1-})},\label{24b}~
	\eea
	\ees
	where $\kappa_1=-\frac{(p_1-il)}{(p_1^*+il)}$,  $\phi_S^{1-}=\frac{1}{2}\log\frac{-i(i\beta+2p_1^*)}{(p_1+p_1^*)}$ and $\phi_L^{1-}=\frac{1}{2}\log\frac{\lvert i\beta+2p_1^*\rvert^2}{(p_1+p_1^*)^2}$. In the latter, superscript $(1-)$ denotes the soliton 1 before collision and subscripts $S$ and $L$ represent the SW and LW, respectively. \\
	Soliton 2: $\eta_{2R}\simeq 0$, $\eta_{1R}\rightarrow +\infty$ 
	\bes
	\bea
	&&S(x,t)=\frac{\tau}{2}e^{i\theta+\Theta_1}\bigg[(1+\kappa_2)-(1-\kappa_2)\tanh(\eta_{2R}+\phi_{S}^{2-})\bigg],\label{25a}\\
	&&L(x,t)=- \frac{4p_{2R}^2}{(\beta-2p_{2I})+\lvert 2p_{2R}+i(\beta-2p_{2I})\rvert \cosh(2\eta_{2R}+\phi_{L}^{2-})}.\label{25b}~
	\eea
	\ees
	Here, 
	\bea
	&&\kappa_2=-\frac{(p_2-il)}{(p_2^*+il)},~~~~~\Theta_1=\log\frac{-(p_1-il)}{p_1^*+il},\nonumber\\&&\phi_S^{1-}=\frac{1}{2}\log\frac{-i\lvert p_1-p_2\rvert^2(i\beta+2p_2^*)}{\lvert p_1+p_2^*\rvert^2(p_2+p_2^*)},~ \text{and}~~
	\phi_L^{1-}=\frac{1}{2}\log\frac{\lvert p_1-p_2\rvert^2\lvert i\beta+2p_2^*\rvert^2}{(p_2+p_2^*)\lvert p_1+p_2^*\rvert^2}.\nonumber
	\eea
	In the above, the superscript $(2-)$ denotes the soliton 2 before collision. \\
	(b) After collision: $t\rightarrow +\infty$\\
	Soliton 1: $\eta_{1R}\simeq 0$, $\eta_{2R}\rightarrow +\infty$ 
	\bes
	\bea
	&&S(x,t)=\frac{\tau}{2}e^{i\theta+\Theta_2}\bigg[(1+\kappa_1)-(1-\kappa_1)\tanh(\eta_{1R}+\phi_{S}^{1+})\bigg],\label{26a}\\
	&&L(x,t)=- \frac{4p_{1R}^2}{(\beta-2p_{1I})+\lvert 2p_{1R}+i(\beta-2p_{1I})\rvert \cosh(2\eta_{1R}+\phi_{L}^{1+})},\label{26b}~
	\eea
	\ees
	where 
	\bea
	&&\Theta_2=\log\frac{-(p_2-il)}{p_2^*+il},~~~ \phi_S^{1+}=\frac{1}{2}\log\frac{-i\lvert p_1-p_2\rvert^2(i\beta+2p_1^*)}{\lvert p_1+p_2^*\rvert^2(p_1+p_1^*)},\nonumber\\ \text{and} &&\phi_L^{1+}=\frac{1}{2}\log\frac{\lvert p_1-p_2\rvert^2\lvert i\beta+2p_1^*\rvert^2}{(p_1+p_1^*)\lvert p_1+p_2^*\rvert^2}.
	\eea
	In the above, the superscript $(1+)$ denotes the soliton 1 after collision.\\
	\underline{Soliton 2}: $\eta_{2R}\simeq 0$, $\eta_{1R}\rightarrow -\infty$ 
	\bes
	\bea
	&&S(x,t)=\frac{\tau}{2}e^{i\theta}\bigg[(1+\kappa_2)-(1-\kappa_2)\tanh(\eta_{2R}+\phi_{S}^{2+})\bigg],\label{27a}\\
	&&L(x,t)=- \frac{4p_{2R}^2}{(\beta-2p_{2I})+\lvert 2p_{2R}+i(\beta-2p_{2I})\rvert \cosh(2\eta_{2R}+\phi_{L}^{2+})}.\label{27b}~
	\eea
	\ees
	In the above, $\phi_S^{2+}=\frac{1}{2}\log\frac{-i(i\beta+2p_2^*)}{p_2+p_2^*}$, $\phi_L^{2+}=\frac{1}{2}\log\frac{\lvert i\beta+2p_2^*\rvert^2}{(p_2+p_2^*)^2}$. Here, the superscript $(2+)$ represents the soliton 2 after collision. 
	\begin{figure*}[]
		\centering
		\includegraphics[width=0.75\linewidth]{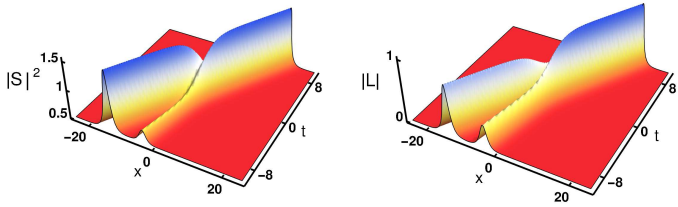}	
		\caption{Elastic collision dynamics of the two anti-dark solitons is displayed with the parameter values $p_1=0.51+0.75i$, $p_2=0.66+0.25i$, $l=1$, $\tau=0.5+0.5i$, and $\beta=1$.  }
		\label{f10}
	\end{figure*}
	\begin{figure*}[]
		\centering
		\includegraphics[width=0.75\linewidth]{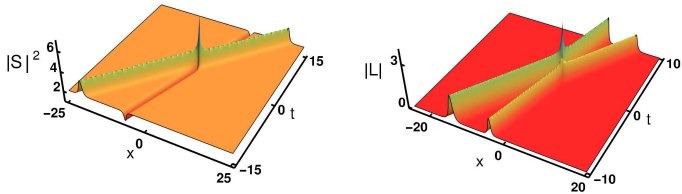}	
		\caption{Elastic collision dynamics between a dark soliton and an anti-dark soliton. }
		\label{f11}
	\end{figure*}

\begin{figure*}[]
		\centering
		\includegraphics[width=0.75\linewidth]{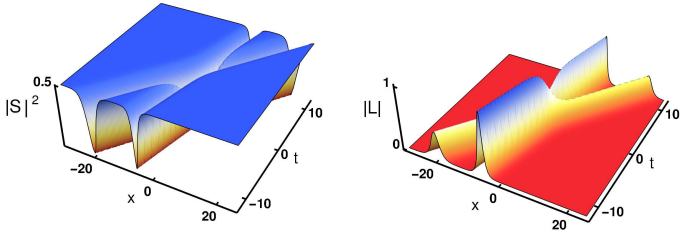}	
		\caption{Collision dynamics of two dark solitons is drawn with the values $p_1=-0.51+0.8i$, $p_2=-0.66+0.25i$, $l=0.85$, $\tau=1+i$, and $\beta=1$.}
		\label{f12}
	\end{figure*}

	The above asymptotic analysis clearly shows that the two anti-dark solitons retain their shape during the collision scenario, except for a finite phase shift, thereby confirming  the elastic nature of the collision. A typical elastic collision among the two anti-dark solitons is depicted in Fig. \ref{f10}. Then, in Fig. \ref{f11} we display the collision between a dark soliton and an anti-dark solitons. To bring out this figure we fix the parameter values as $p_1=1+0.65i$, $p_2=-1.5+0.25i$, $l=0.85$, $\tau=1+i$, and $\beta=1$. From  Fig. \ref{f11}, it is evident that the dark and anti-dark solitons are well separated initially and their structures are invariant under collision.  A similar situation is also observed during the interaction among the two dark solitons and this scenario is depicted in Fig. \ref{f12}.  We have calculated the phase shift suffered by the two anti-dark solitons during the collision process and they turn out to be
	\begin{equation}
		\Del\Phi_{SW}^1=\frac{1}{2}\log\frac{\lvert p_1-p_2\rvert ^2}{\lvert p_1+p_2^*\rvert ^2}=-\Del\Phi_{SW}^2,~\Del\Phi_{LW}^1=-\Del\Phi_{LW}^2=2\Phi_{SW}^1.\label{28}
	\end{equation} 
	\begin{figure*}[]
		\centering
		\includegraphics[width=0.8\linewidth]{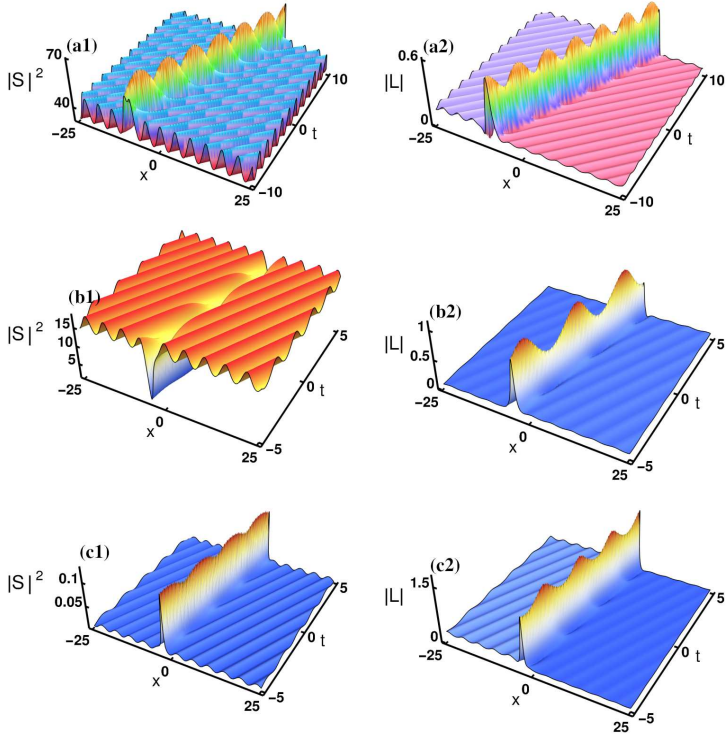}	
		\caption{Dark solitons behave like a breather in the periodic background field. In Figs. (a1) and (c1) we display bright breather like behaviour of anti-dark soliton in the SW component whereas in Fig. (b1) dark breather like pattern is observed in the SW component. In contrast to this, in all the figures (a2), (b2) and (c2), a bright breather like pattern is observed in the LW component.  }
		\label{f13}
	\end{figure*}
	\begin{figure*}[]
		\centering
		\includegraphics[width=0.85\linewidth]{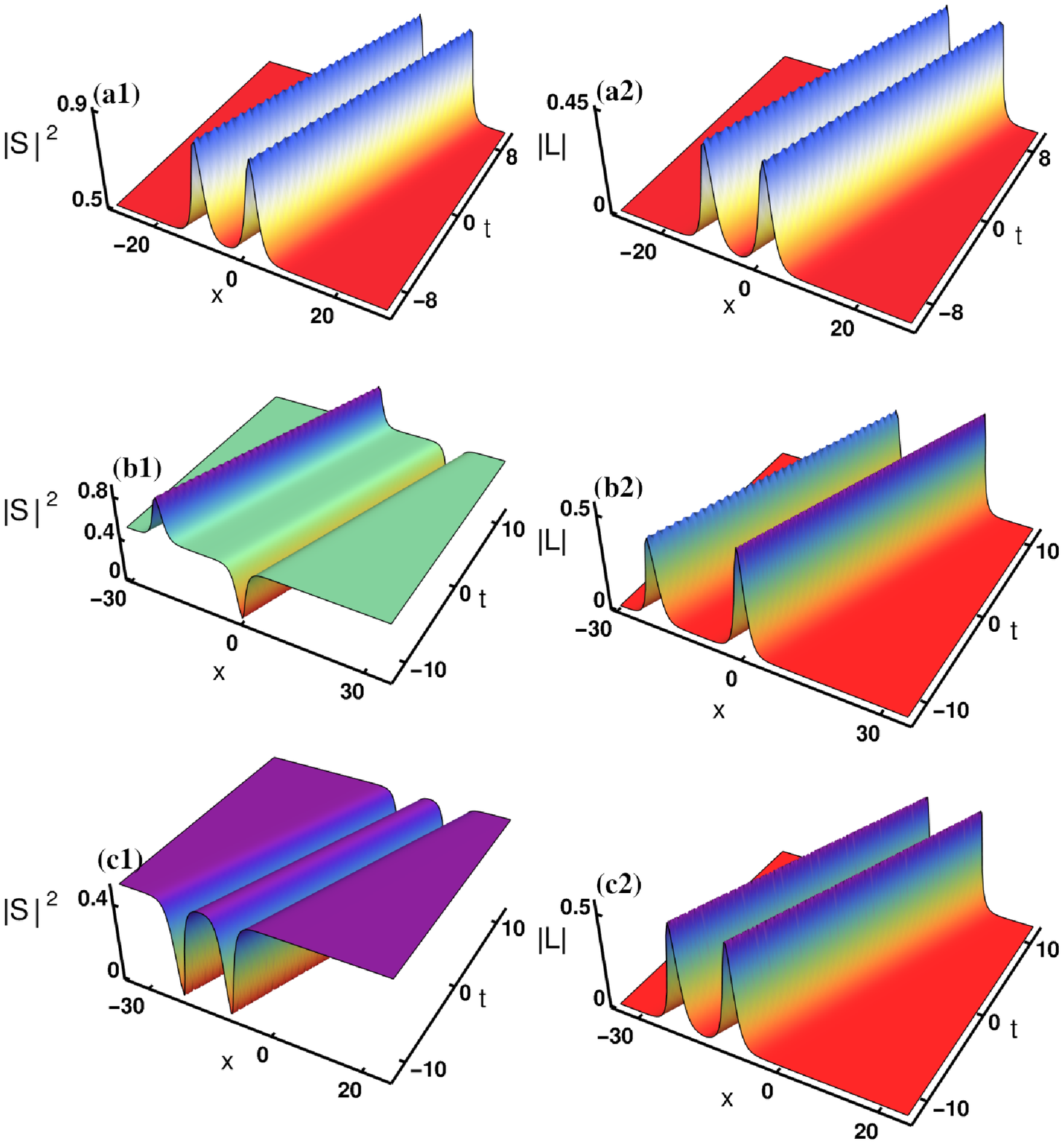}	
		\caption{The different types of dark soliton bound states are demonstrated. In Fig (a1), we display anti-dark soltion bound state structure whereas in Fig. (b1) we illustrate the existence of anti-dark-dark solitons  bound state. Then, two dark-solitons bound state structure is depicted in Fig. (c1). In addition to these bound state structure of the SW component,  a parallel propagating bright-solitons bound structure always appears in the LW component and they are illustrated in Figs. (a2), (b2) and (c2).  The parameter values are: (i) (a1)-(a2):   $p_1=0.5+0.45i$, $p_2=0.5+0.4445i$, $l=1$, $\tau=0.5+0.5i$, and $\beta=1$. (ii) (b1)-(b2): $p_1=-0.36+0.515i$, $p_2=0.36+0.5i$, $l=0.65$, $\tau=0.5+0.5i$, and $\beta=1$. (iii) (c1)-(c2): $p_1=-0.5+0.45i$, $p_2=-0.5+0.4445i$, $l=1$, $\tau=0.5+0.5i$, and $\beta=1$.   }
		\label{f14}
	\end{figure*}
	As we pointed in the one-dark soliton case, the two-dark soliton solution also exhibits periodic behaviour for  low values of the wave number $l$ of the background wave $\tau e^{i\theta}$. Such a possibility is illustrated in Fig. \ref{f13}. From this figure, one can identify that the two dark solitons do not completely change into periodic waves. On the other hand, one of the dark/anti-dark solitons
	behave like a breather in a periodic background wave field. From Figs. \ref{f13}(a1)-(b1), we observe that an anti-dark soliton (or a dark-soliton) in the SW component turns into a bright breather (or dark breather) like structure on the periodic wave background. From Figs. \ref{f13}(a2)-(b2), we also observe a bright breather-like pattern in the LW component. In addition to this, a breathing pattern is observed in both the SW and LW components, which is demonstrated in Figs. (\ref{f13}) (c1)-(c2). The presence of a dark soliton in the periodic background will be useful in connection with the recent literature on the theory of rogue waves in periodic background wave field \cite{psky1,psky2}. Further, in Fig. \ref{f14}, we display the three types of parallelly propagating dark-soliton bound states. We note that the resonance soliton and breathing type bound state do not exist in the dark-soliton case.    
	\begin{figure*}[]
		\centering
		\includegraphics[width=0.85\linewidth]{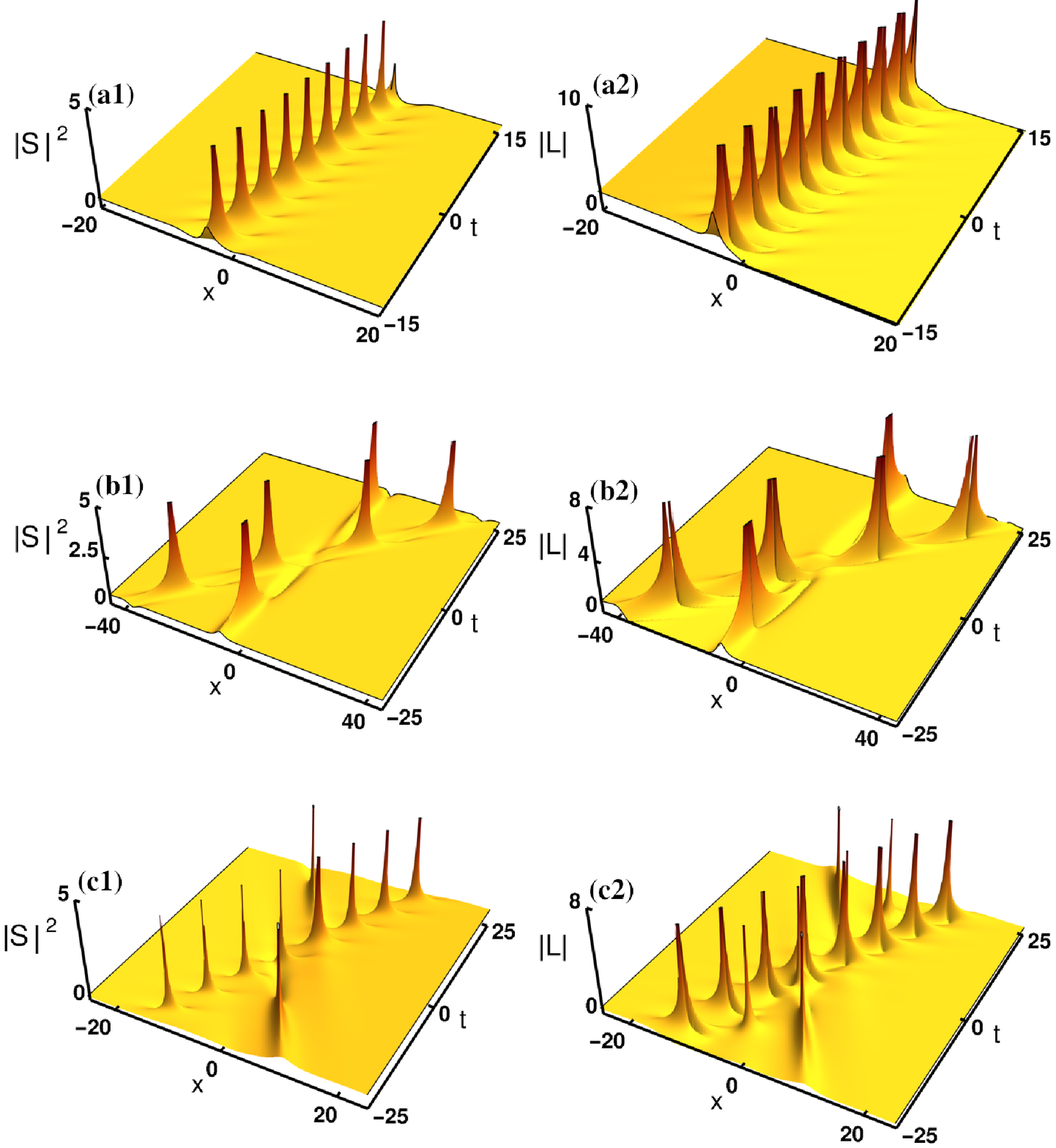}	
		\caption{Breather solution of the generalized LSRI system (\ref{1}) is illustrated for $\beta>0$. The parameter values are fixed as follows: (a1)-(a2): $\phi_1=0.5i$, $\phi_2=\phi_1^*+\pi$, $l=0.5$, $\tau=0.5$, and $\beta=1$. (b1)-(b2): $\phi_1=0.75$, $\phi_2=\phi_1^*+\pi$, $l=0.5$, $\tau=0.5$, and $\beta=0.25$. (c1)-(c2): $\phi_1=0.25+0.25i$, $\phi_2=\phi_1^*+\pi$, $l=0.5$, $\tau=0.5$, and $\beta=0.25$.  }
		\label{f15}
	\end{figure*}
	\begin{figure*}[]
		\centering
		\includegraphics[width=0.85\linewidth]{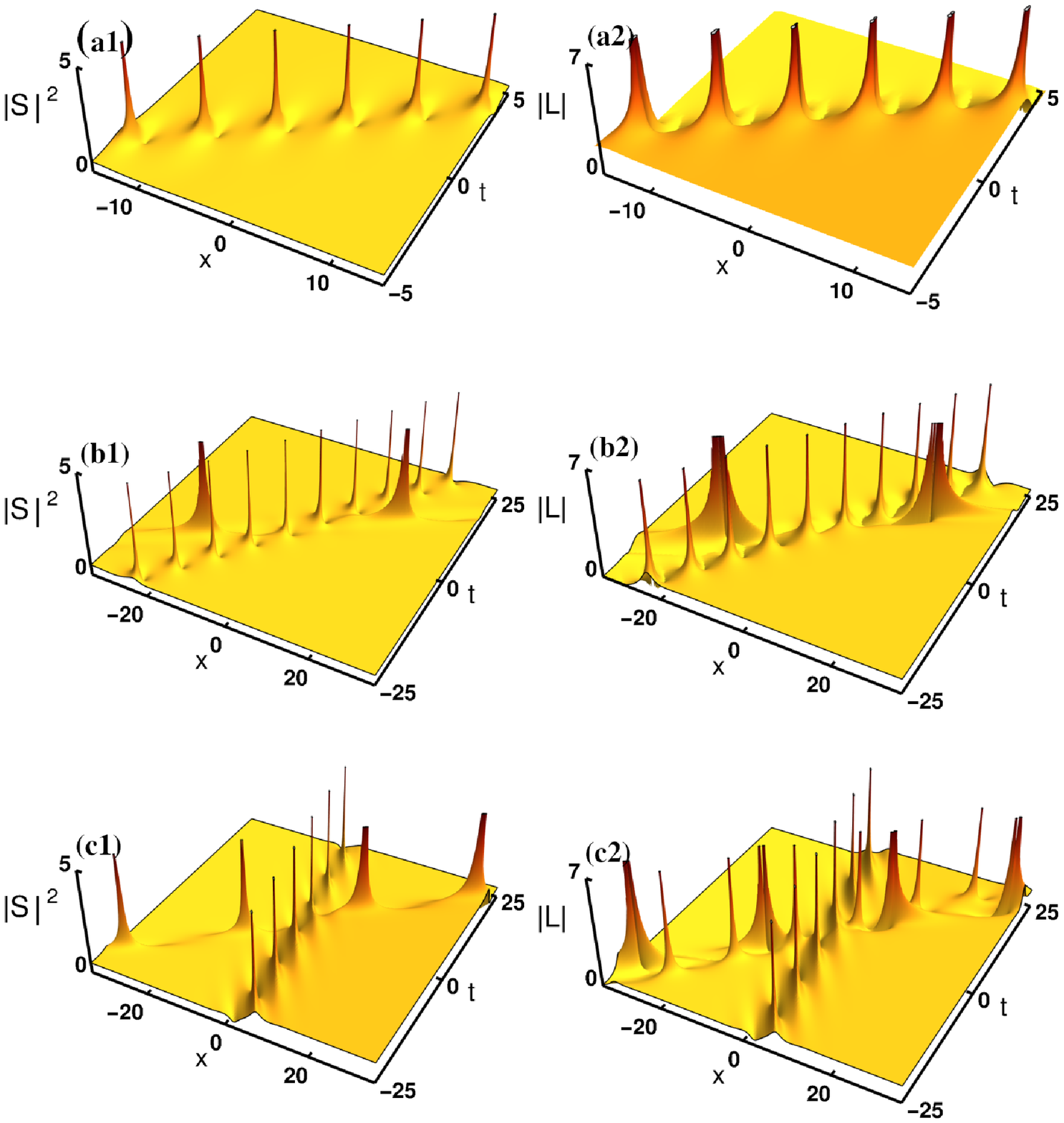}	
		\caption{Breather solution of the generalized LSRI system (\ref{1}) is illustrated for $\beta<0$.  A singular breather periodic in both $x$ and $t$ is demonstrated in Figs. (a1)-(a2) for $\phi_1=0.5$, $\phi_2=\phi_1^*+\pi$, $l=0.5$, $\tau=0.5$, and $\beta=-1$. The two interacting breathers are illustrated in Figs. (b1)-(b2) for $\phi_1=0.35$, $\phi_2=\phi_1^*+\pi$, $l=0.5$, $\tau=0.5$, and $\beta=-0.25$. In Fig. (c1), we find that a stationary breather and a moving breather are emerging out from the SW component. In contrast to this,  in the LW component, a stationary breather along with the two interacting breathers are observed. To display Figs. (c1)-(c2), we set the parameter values as $\phi_1=0.35+0.35i$, $\phi_2=\phi_1^*+\pi$, $l=0.5$, $\tau=0.5$, and $\beta=-0.25$.    }
		\label{f16}
	\end{figure*}
	\section{Breather solution}
	To get the breather solution, one has to consider the same bilinear transformation (Eq. (\ref{14})) that has been used to derive the dark-soliton solution. By doing so, we obtain the following
 functions 	$g$ and $f$ corresponding to the breather solution of the generalized LSRI system (\ref{1}):
\bes\begin{eqnarray}
		&&\hspace{-1.0cm}g=1+e^{\eta_1+2i\phi_1}+e^{\eta_2+2i\phi_2}+A_{12}e^{\eta_1+\eta_2+2i(\phi_1+\phi_2)},\label{30a}\\
		&&\hspace{-1.0cm}f=1+e^{\eta_1}+e^{\eta_2}+A_{12}e^{\eta_1+\eta_2},\label{30b}\\
		&&\hspace{-1.0cm}\text{where}\nonumber\\&&\hspace{-1.0cm}	A_{12}=\frac{1}{D}\bigg(p_1^2\sin(\phi_1-\phi_2)[\sin(\phi_1+\phi_2)-\sin(\phi_1-\phi_2)]-p_2^2\sin(\phi_1-\phi_2)\nonumber\\
		&&\times[\sin(\phi_1+\phi_2)+\sin(\phi_1-\phi_2)]-(p_1-p_2)^2\cos(\phi_1-\phi_2)\nonumber\\
		&&\times[\cos(\phi_1-\phi_2)-\cos(\phi_1+\phi_2)]\bigg),\nonumber\\
		&&\hspace{-1.0cm}D=-p_1^2\sin(\phi_1+\phi_2)[\sin(\phi_1+\phi_2)-\sin(\phi_1-\phi_2)]-p_2^2\sin(\phi_1+\phi_2)\nonumber\\
		&&\times[\sin(\phi_1+\phi_2)+\sin(\phi_1-\phi_2)]+(p_1+p_2)^2\cos(\phi_1+\phi_2)\nonumber\\
		&&\times[\cos(\phi_1-\phi_2)-\cos(\phi_1+\phi_2)],\nonumber
	\end{eqnarray}\ees
	where $\eta_j=p_jx-\Omega_jt$, $\Omega_j=2lp_j-p_j^2\cot\phi_j$, $p_1=\frac{1}{2}\bigg(i\beta-\sqrt{-\beta^2+16\lvert\tau\vert^2\sin^2\phi_1+8i\lvert \tau\rvert^2\sin2\phi_1}\bigg)$, $p_2=\frac{1}{2}\bigg(i\beta+\sqrt{-\beta^2+16\lvert\tau\vert^2\sin^2\phi_2+8i\lvert \tau\rvert^2\sin2\phi_2}\bigg)$. Here, $p_j$, $\Omega_j$ and $\phi_j$, $j=1,2,$ are complex constants.  A typical singular time-periodic breather is displayed in Fig. \ref{f15}(a1)-(a2)  with the parameter values  $\phi_1=0.5i$, $\phi_2=\phi_1^*+\pi$, $l=0.5$, $\tau=0.5$, and $\beta=1$. This figure shows that the breather obtained by us is similar to Kuznetsov-Ma soliton \cite{kma} which has been widely discussed  in the context of rogue-waves. For $\beta=0.25$, the solution (\ref{30a})-(\ref{30b}) admits two interacting breathers in both the components, where one of the breathers is stationary along $x=0$. This is illustrated in Fig. \ref{f15}(b1)-(b2). Further, as we demonstrated in Fig. \ref{f15}(c1)-(c2), we also come across another breather pattern by considering the phase, $\phi_1$, as complex and for a low positive value of $\beta$. From this pattern, we observe that, in the SW component, the two breathers propagate in opposite directions and they collide with each other. The final outcome is reflected in changing their positions. In contrast to this, in the LW component, in addition to the two interacting breathers moving in the opposite directions, there is a stationary breather that appears along $x=0$. Furthermore, one also gets similar breather patterns for $\beta<0$. Such a possibility is illustrated in Fig. \ref{f16}.        
	\section{Conclusion}\label{sec13}
	In this paper, first we have derived $N$-bright and $N$-dark soliton solutions for the generalized LSRI system (\ref{1}) through the Hirota bilinear method. Then, by considering the fundamental bright and dark soliton solutions as well as their higher-order forms, we have discussed their various propagation and collision properties in detail. The interesting aspect of the present generalized LSRI system is that the bright soliton, in general, behaves like  KdV soliton. However, under a special condition, it acts like the NLS soliton. Further, we found that the dark-soliton admits three types of dark soliton profiles. Further, the asymptotic analysis confirmed that both the bright and dark solitons always exhibit elastic collision only. In addition to these, we also demonstrated the existence of resonant interactions among the two bright solitons, and soliton molecules. Finally, by deriving the breather solution  we have illustrated the various breather patterns graphically by tuning the phase values and a system parameter $\beta$. The present study will be useful in fluid dynamics, plasma physics, nonlinear optics and other closely related disciplines of physics.

	\backmatter
	

	\bmhead{Acknowledgments}
	
	The works of Mokhtar Kirane, and Stalin Seenimuthu, are supported by Khalifa University of Science and Technology, Abu-Dhabi, UAE, under the Project Grant No. 8474000355. Lakshmanan Muthusamy thanks DST-SERB for the award of a DST-SERB National Science Chair (NSC/2020/000029). 
	
	\section*{Declarations}

	\begin{itemize}
		\item 
		The authors declare that they have no conflict of interest
		\item All data generated or analyzed during this
		study are included in the article 
	\end{itemize}
	
	
	
	
	
	
	
	\begin{appendices}
		
		\section{$N$-bright soliton solution}\label{secA1}
		
		The explicit form of $N$-bright soliton solution of Eq. (\ref{1}) can be expressed using Gram determinant in the following way:
		\begin{eqnarray}
			g=\begin{vmatrix}
				A & I & \phi^T \\ 
				-I &B & {\bf 0}^T \\ 
				{\bf 0}  & -C & 0
			\end{vmatrix},~~f=\begin{vmatrix}
				A & I \\ 
				-I &B \\ 
			\end{vmatrix},~~f^*=\begin{vmatrix}
				A' & I \\ 
				-I &B^* \\ 
			\end{vmatrix},\label{A.1a}
		\end{eqnarray}
		The various elements of matrices $A$, $A'$ and $B$ are obtained from the following, 
		\begin{eqnarray}
			A_{ij}=\frac{k_j^*}{(k_i+k_{j}^*)}e^{\eta_i+\eta_{j}^*},~ A_{ij}'=-\frac{k_i}{(k_i+k_{j}^*)}e^{\eta_i+\eta_{j}^*},b_{ij}=-\frac{\ga_i^*\ga_j(i\beta+2k_{j}^*)}{(k_i^{*2}-k_j^{2})},\nonumber
		\end{eqnarray}
		$\eta_j=k_jx+ik_j^2t$, $i, j=1,2,...,N$.
		The row matrices in Eq. (\ref{A.1a}) are defined below: \\
		$\phi=\begin{pmatrix}
			e^{\eta_1} & e^{\eta_2}  & .  & .  & . & e^{\eta_N}
		\end{pmatrix}$, $C=\begin{pmatrix}
			\ga_1 & \ga_2 & . & . & . &\ga_N
		\end{pmatrix}$, ${\bf 0}$ is a $N$-component zero row matrix and $\sigma=I$ is a $(N\times N)$ identity matrix. The above $N$-soliton solution is characterized by $(2N)$ arbitrary complex parameters, $k_j$ and $\ga_j$, $j=1,2$ and one system parameter $\beta$.
		\section{$N$-dark soliton solution}\label{secA2}
		The $N$-sark soliton solution of the system (\ref{f1})
		is given by 
		\bes\begin{eqnarray}
			&&g=\begin{vmatrix}\displaystyle{
					\delta_{jk}+i\bigg(\frac{i\beta+2p_k^*}{p_j+p_k^*}}\bigg)\bigg(\frac{p_j-il}{p_j^*+il}\bigg)e^{\eta_j+\eta_k^*}
			\end{vmatrix}_{N\times N},\\
			&&f=\begin{vmatrix}\displaystyle{
					\delta_{jk}-i\bigg(\frac{i\beta+2p_k^*}{p_j+p_k^*}}\bigg)e^{\eta_j+\eta_k^*}
			\end{vmatrix}_{N\times N},
		\end{eqnarray}\ees
		where $\eta_j=p_jx+ip_j^2t+\eta_{j}^{(0)}$, $j=1,2,...,N$, $p_j$'s and $\eta_j^{(0)}$'s are complex constants. The constraint conditions are obtained and they turn out to be  
		$p_{jR}=\pm \bigg[\frac{\lvert\tau\rvert^2(2l-\beta)}{2p_{jI}}-(p_{jI}-l)^2\bigg]^{\frac{1}{2}}$, ~$j=1,2,...,N$. Here, $p_{jR}$ and $p_{jI}$'s are the real and imaginary parts of $p_j$'s. The imaginary parts of $p_j$'s govern the velocity of the solitons and $\eta_{j}^{(0)}$'s define the phase of the solitons.  
		
		
		
	\end{appendices}
	
	
	
	
\end{document}